\documentclass[%
 aip,
 amsmath,amssymb,
 reprint,%
]{revtex4-1}

\usepackage{amsmath, amssymb, accents, float} % note that this was added

\usepackage{graphicx}% Include figure files
\usepackage{dcolumn}% Align table columns on decimal point
\usepackage{bm}% bold math
\usepackage[utf8]{inputenc}
\usepackage[T1]{fontenc}
\usepackage{mathptmx}
\usepackage{etoolbox}

\makeatletter
\def\@email#1#2{%
 \endgroup
 \patchcmd{\titleblock@produce}
  {\frontmatter@RRAPformat}
  {\frontmatter@RRAPformat{\produce@RRAP{*#1\href{mailto:#2}{#2}}}\frontmatter@RRAPformat}
  {}{}
}%
\makeatother
\begin{document}

\preprint{AIP/123-QED}

\title[$N_1$ in LAOS]{Self-Consistent Fourier-Tschebyshev Representations of the First Normal Stress Difference in Large Amplitude Oscillatory Shear}
% Force line breaks with \\
\author{Nicholas King}
\affiliation{ 
Department of Chemical Engineering, Massachusetts Institute of Technology%\\This line break forced with \textbackslash\textbackslash
}%

\author{Eugene Pashkovski}%
\author{Reid Patterson}
\author{Paige Rockwell}
\affiliation{ 
Lubrizol Corporation, Research and Development%\\This line break forced with \textbackslash\textbackslash
}%

\author{Gareth H. McKinley *}
\email{gareth@mit.edu}
\affiliation{%
Department of Mechanical Engineering, Massachusetts Institute of Technology%\\This line break forced% with \\
}%

\date{\today}% It is always \today, today,

\begin{abstract}
Large Amplitude Oscillatory Shear (LAOS) is a key technique for characterizing nonlinear viscoelasticity in a wide range of materials. Most research to date has focused on the shear stress response to an oscillatory strain input. However, for highly elastic materials such as polymer melts, the time-varying first normal stress difference $N_1(t;\omega,\gamma_0)$ can become much larger than the shear stress at sufficiently large strains, serving as a sensitive probe of the material's nonlinear characteristics. We present a Fourier-Tschebyshev framework for decomposing the higher-order spectral content of the $N_1$ material functions generated in LAOS. This new decomposition is first illustrated through analysis of the second-order and fourth-order responses of the quasilinear Upper Convected Maxwell model and the fully nonlinear Giesekus model. We then use this new framework to analyze experimental data on a viscoelastic silicone polymer and a thermoplastic polyurethane melt. Furthermore, we couple this decomposition with the recently developed Gaborheometry strain sweep technique to enable rapid and quantitative determination of the $N_1$ material function from experimental normal force data obtained in a single sweep from small to large strain amplitudes. We verify that asymptotic connections between the oscillatory shear stress and $N_1$ in the quasilinear limit are satisfied for the experimental data, ensuring self-consistency. This framework for analyzing the first normal stress difference is complementary to the established framework for analyzing the shear stresses in LAOS, and augments the content of material-specific data sets, hence more fully quantifying the important nonlinear viscoelastic properties of a wide range of soft materials.
\end{abstract}

\maketitle

\section{Introduction: Large Amplitude Oscillatory Shear}

Large Amplitude Oscillatory Shear (LAOS) has been well-established as a critical tool for measuring the nonlinear properties of many viscoelastic fluids, ranging from polymeric liquids \cite{hyun_review_2011} to food products \cite{wang_food_2022} and many other material samples encountered in areas of soft matter. The resultant set of material parameters obtained from analysis of the data serves as a detailed \textit{rheological fingerprint} of a fluid useful for quality specifications and optimization of industrial processes. In a typical commercial rheometer, a small quantity of fluid is sandwiched between a cone and a plate (or any other specific geometry), and the fluid is sheared between the two surfaces. For an oscillatory input, the amplitude and frequency of the applied strain control the amplitude and frequency of the output stress signal. In the limit of small strain amplitude (also known as Small Amplitude Oscillatory Shear (SAOS)), the stress response is linear, and the results can be interpreted within the well-established framework of linear viscoelasticity. However, for large strain amplitudes (LAOS), the fluid response becomes nonlinear, and the shear stress amplitude varies nonlinearly with time and the applied strain amplitude. Oscillatory normal stress differences also start to become significant. By their symmetry, these two independently measured signals can be decomposed respectively into a Fourier series of odd harmonics for the shear stress and even harmonics for the normal stresses. Such an analysis of LAOS data is known as Fourier Transform (FT) rheology \cite{wilhelm_fourier-transform_2002}. Other methods of analyzing the signal include the stress decomposition technique of Cho \textit{et al. }\cite{cho_geometrical_2005}, the widely-used Fourier-Tschebyshev decomposition by Ewoldt \textit{et al.} \cite{ewoldt_new_2008}, and the Sequence of Physical Processes (SPP) procedure proposed by Rogers and coworkers \cite{rogers_search_2017}. 

However, almost all studies of LAOS have focused exclusively on the shear stress response, even though oscillatory normal stress differences also become significant at large strain amplitudes. Normal stresses resulting from shearing deformations are common in solid mechanics, with well-known examples such as the Poynting effect, in which a cylinder extends axially when subject to torsion due to the normal stresses that develop \cite{anand_continuum_2020}. In soft solids such as brain tissue, the normal stresses arising during the rapid shearing displacements experienced in vehicle accidents can contribute to traumatic brain injury \cite{balbi_poynting_2019}. In the polymer processing industry, these normal stress differences can dominate in the strong flow fields that are routinely encountered in manufacturing processes such as film stretching and fiber spinning. However, the detailed functional form of the first normal stress differences generated in LAOS, and how they evolve with strain amplitude, has been little explored \cite{nam_analysis_2008, nam_first_2010}, and there is no standard nomenclature for referring to the material coefficients that characterize each harmonic term. Furthermore, as an intrinsically nonlinear effect, the evolution of $N_1$ may be expected to be a more sensitive probe of the rheological characteristics (fingerprint) of an unknown viscoelastic material, and therefore might prove especially useful for discovering or validating the predictions of nonlinear constitutive equations specific to a material \cite{hyun_review_2011}. 

Measures of the first normal stress difference have been refined over many years. Philippoff's accurate measurements of $N_1$ in the 1950s \cite{philippoff_normal_1957} subsequently played a role in his nomination for the Bingham medal. Many more advances in normal stress measurements have been made in recent years and are summarized in a recent review \cite{costanzo_review_2024}. The second normal stress difference ($N_2$) in shear is commonly much smaller than the first normal stress difference ($N_1$) in polymeric liquids, and is also more difficult to measure accurately \cite{maklad_review_2021}. Due to the complexities of the measurements, a wide range of different geometries on commercial rheometers has been developed. In particular, the cone and partitioned plate geometry has been helpful in delaying the onset of edge fracture \cite{li_normal_2025}, and we consider this geometry in Section IV below. 

One interesting feature of $N_1$ in oscillatory shear that results from the general memory integral expansion is that there should be an internal "self-consistency" between the $G'$ and $G''$ values determined from the measured shear stress and the zeroth and second harmonics of $N_1$ measured at low to moderate strain amplitudes $\gamma_0$. Such consistency relations were first explored in the 1960s by Williams and Bird \cite{williams_oscillatory_1964} for the Upper Convected Maxwell model and more generally by Coleman and Markovitz \cite{coleman_asymptotic_1974} for the "simple fluid". Experimentally, these relationships have also been demonstrated for a number of polymeric liquids. Endo and Nagasawa \cite{endo_normal_1970} performed frequency-sweep oscillatory shear experiments with polystyrene dissolved in a chlorinated diphenyl solution and compared the magnitudes of the zeroth harmonic (i.e. the `DC' offset) of $N_1$ with the values of $G'$ obtained from the oscillating shear stress in the fluid, obtaining good agreement between the two. However, they reported poor agreement when comparing the magnitude of the second harmonics of $N_1$ with the corresponding values computed from the nonlinear shear stress data. Christiansen and Leppard \cite{christiansen_steady-state_1974} performed frequency-sweep oscillatory shear experiments with polyethylene oxide (PEO) and polyacrylamide (PAA) solutions, and obtained very good agreement between the zeroth and second harmonics of $N_1$, and the corresponding shear stress data. In both of these papers, moderate strain amplitudes of less than 100\% were chosen, and the data was compared across various angular frequencies. Subsequently, Isayev and Hieber \cite{isayev_oscillatory_1982} used strain-sweep oscillatory shear experimental data on polybutadiene melts and plotted the evolution of the zeroth harmonic of $N_1$ against $\gamma_0^2$, comparing it with predictions of the nonlinear viscoelastic model proposed by Leonov \cite{leonov_nonequilibrium_1976,leonov_theoretical_1976}. These experiments did not reach low enough values of the strain amplitude to definitively confirm agreement between the $N_1$ and shear stress data. Separately, starting from the simple fluid model of Coleman and Noll \cite{coleman_foundations_1961}, Vrentas \cite{vrentas_finite_1991} calculated the leading-order asymptotic nonlinearities of the shear stress and $N_1$ for a steady shearing flow from LAOS data. 

Apart from these specific works from decades ago, the measurement of $N_1$ in LAOS has not gained much traction within the rheology community. The works of Endo and Nagasawa \cite{endo_normal_1970}, Christiansen and Leppard \cite{christiansen_steady-state_1974} and Isayev and Hieber \cite{isayev_oscillatory_1982} also primarily focused on demonstrating the self-consistency of the normal stress measurements with shear stress measurements at low and moderate $\gamma_0$ (when second order terms $\mathcal{O}(\gamma_0^2)$ dominate), and they did not systematically extend their experiments to larger values of $\gamma_0$, in which higher-order even harmonics also appear in the $N_1$ signal. Each of these studies also used vastly different terminology to refer to the material coefficients that control the amplitude of the higher-order even harmonics of the normal stress response. In light of the popularity of the Fourier-Tschebyshev polynomial representation of shear stress in LAOS, we seek to create a unifying notation for $N_1$ in LAOS and to show experimentally using a modern separate motor transducer (SMT) rheometer \cite{costanzo_review_2024} that self-consistency between the zeroth and second-order harmonics of the $N_1$ signal and the first harmonics of the shear stress signal is satisfied. 

Another benefit of LAOS rheometry is the richness and fidelity of the datasets that can be obtained. There has been a dramatic growth in data-driven modeling across the field of rheology in recent years \cite{mangal_data-driven_2025}, and normal stress measurements as a function of frequency and strain amplitude amount to an extended rheological fingerprint of a complex fluid. A framework for understanding and extracting information from the $N_1$ response of a viscoelastic fluid in LAOS will also be useful as a feature representation for machine learning techniques. We therefore aim to clarify the notation for the key measurable quantities and to build upon recent experimental innovations that have been applied to analysis of the time- and strain- dependence of the shear stress measured in LAOS, such as Gaborheometry \cite{rathinaraj_gaborheometry_2023}, which enables faster acquisition of data (especially important for time-varying or mutating materials), across different amplitudes and frequencies of the applied strain.

The paper is organized as follows. We first introduce the Fourier-Tschebyshev framework for analyzing the structure of $N_1$ in LAOS, using the Upper Convected Maxwell model as an illustrative example. We next analyze the Giesekus model, a fully nonlinear model, showing that it both satisfies self-consistency and also produces higher harmonics. We then use experimental data for a relatively simple polydimethylsiloxane (PDMS) silicone fluid and a more rheologically complex thermoplastic polyurethane (TPU) fluid to illustrate these effects in real materials. 

\section{Theory: Development of a framework for quantifying material functions of $N_1$ in LAOS}

In the Fourier-Tschebyshev formulation \cite{ewoldt_new_2008}, the shear stress $\sigma$ measured in LAOS, driven by an applied strain with amplitude $\gamma_0$ and angular frequency $\omega$, is expanded in terms of Tschebyshev polynomials $T_k$:
\begin{equation}
\begin{split}
    \sigma &= \gamma_0 \sum_{k=1,\text{ odd}}^{\infty} \left( G_k' \sin (k\omega t) + G_k'' \cos (k\omega t) \right) \\
    &= \gamma_0 \sum_{k=1, \text{ odd}}^{\infty} e_k T_k (x) + \gamma_0 \omega \sum_{k=1,\text{ odd}}^{\infty} v_k T_k (y)
\end{split}
\end{equation}
This decomposition is based on the fact that the applied strain and strain rate are orthogonal to each other, and can be expressed in terms of the normalized strain $x = \gamma / \gamma_0 = \sin(\omega t)$ and the normalized strain rate $y = \dot{\gamma} / (\gamma_0 \omega) = \cos (\omega t)$. Ewoldt \textit{et al.} \cite{ewoldt_new_2008} defined the coefficients weighting each spectral component as elastic coefficients $e_k$ and viscous coefficients $v_k$. This has enabled more physical interpretations of common terms such as strain-softening or hardening and shear-thinning or thickening. 

In the linear regime of low strain amplitudes $\gamma_0 \ll 1$, $e_1$ reduces to the storage modulus $G'$ and $v_1 \omega$ reduces to the loss modulus $G''$, as per the requirements of linear viscoelasticity. All other higher harmonics vanish asymptotically. However, as the strain amplitude increases, material nonlinearities become important and normal stress differences begin to be generated. A quasilinear constitutive equation, such as the Upper Convected Maxwell or Oldroyd-B model, is linear in the stress, but contains a convected differential operator that ensures frame-indifferent material behavior and produces nonlinearities in response to an imposed deformation at a fundamental frequency $\omega$. At moderate strain amplitudes, when nonlinear effects are weak, only a constant term (the zeroth harmonic or `DC' contribution) and second harmonic terms are present in the $N_1$ signal. The functional form of the oscillating material response of $N_1$ in LAOS can be expressed in two equivalent forms, referenced to either the oscillating applied strain or strain rate \cite{ferry_viscoelastic_1980,nam_first_2010}. This is analogous to how the measured shear stress in LAOS can be expressed in terms of the storage and loss moduli $G_k'$ and $G_k''$, or equivalently, the dynamic viscosities $\eta_k'$ and $\eta_k''$. The strain rate form was suggested as early as 1964 by Williams and Bird \cite{williams_oscillatory_1964}, and later updated to more modern notation by Ferry \cite{ferry_viscoelastic_1980} in terms of three material functions $\Psi_1^d$, $\Psi_1'$ and $\Psi_1''$. At second order, the first normal stress difference is thus written:
\begin{equation}
    N_1 (t; \omega,\gamma_0)= (\gamma_0 \omega) ^2 \left[\Psi_1^d + \Psi_1' \cos(2\omega t) + \Psi_1'' \sin (2\omega t) \right]
\end{equation}
These material functions $\Psi_1^d$, $\Psi_1'$ and $\Psi_1''$ (all with units of $Pa.s^2$) are only a function of frequency $\omega$ in the quasilinear regime, but can also become functions of strain amplitude $\gamma_0$ at larger strain amplitudes.

Nam \textit{et al.}\cite{nam_prediction_2008, nam_first_2010} have proposed an alternate notation in terms of the strain amplitude, with material functions $n_1^d$, $n_1'$ and $n_1''$. These functions are similarly only functions of the frequency $\omega$ in the quasilinear regime, but will become functions of strain amplitude $\gamma_0$ at large strain amplitudes. However, in contrast to their original proposed notation, we here propose that their definition of $n_1'$ and $n_1''$, which are the coefficients for the second-order trigonometric terms, be swapped. The rationale for this will be explained below. We hence propose the following form:
\begin{equation} \label{propose}
    N_1 = \gamma_0^2 \left[ n_1^d + n_1' \cos(2\omega t) + n_1'' \sin (2\omega t) \right]
\end{equation}

With this convention, we thus have $n_1^d = \omega^2 \Psi_1^d$, $n_1' = \omega^2 \Psi_1'$, and $n_1'' = \omega^2 \Psi_1''$. Using this notation, the extension to a Fourier series with higher harmonics can be written in the form:
\begin{equation} \label{definition1}
    N_1 = \gamma_0^2 \left( n_1^d + \sum_{k=2,\text{ even}}^{\infty} n_{1,k}' \cos (k\omega t) + n_{1,k}'' \sin (k\omega t) \right)
\end{equation}

Alternately, we can also represent the oscillating harmonic components at each frequency in terms of an amplitude $|n_{1,k}^*| = \sqrt{(n_{1,k}')^2 + (n_{1,k}'')^2}$ and phase $\Phi_k = \tan^{-1} (n_{1,k}'' / n_{1,k}')$, and this leads to the following expressions:
\begin{subequations} \label{definition}
\begin{align}
\begin{split}
    N_1 &= \gamma_0^2 \left( n_1^d + \sum_{k=2,\text{ even}}^{\infty} |n_{1,k}^*| \cos (k\omega t - \Phi_k) \right)
\end{split} \\
\begin{split} \label{subdefinition}
    N_1 &= \gamma_0^2 \left( n_1^d + \sum_{k=2,\text{ even}}^{\infty} |n_{1,k}^*| \cdot T_k (z) \right)
\end{split}
\end{align}
\end{subequations}

Equation \ref{definition} expresses $N_1$ in a Fourier-Tschebyshev formulation, in which $T_k(z)$ is the $k^{\text{th}}$ Tschebyshev polynomial and $z = \cos (\omega t - \Phi_k/k)$ is a scalar argument that is phase-shifted from the input waveform by a phase $\Phi_k/k$. The use of Tschebyshev polynomials in equation \ref{subdefinition} makes this $N_1$ formulation consistent with the Fourier-Tschebyshev framework commonly used for shear stress in LAOS.

The application of this framework will be illustrated using the Upper Convected Maxwell (UCM) model, which is the simplest frame-indifferent model that describes the rheological characteristics of a polymeric liquid. The UCM model is given by the following expression for the polymer extra stress $\pmb{\sigma}_p$:
\begin{equation}
    \tau \accentset{\triangledown}{\pmb{\sigma}}_p + \pmb{\sigma}_p = G \tau \dot{\pmb{\gamma}}
\end{equation}
The UCM model is parameterized by a modulus $G$ and a single relaxation time $\tau$, where the inverted triangle indicates the upper convected derivative $\accentset{\nabla}{\pmb{\sigma}}_p \equiv \frac{D \pmb{\sigma}_p}{D t} - (\nabla \pmb{v})^T \cdot \pmb{\sigma}_p - \pmb{\sigma}_p \cdot \nabla \pmb{v}$. In oscillatory shear, with strain $\gamma (t) = \gamma_0 \sin (\omega t)$ and strain rate $\dot{\gamma} (t) = \gamma_0 \omega \cos (\omega t)$, an analytical solution for the limit cycle of the dynamical system is possible:
\begin{widetext}
\begin{equation}
    \sigma_{12} = \gamma_0 \left(\frac{G (\omega \tau)^2}{1 + (\omega \tau)^2} \sin (\omega t) + \frac{G (\omega \tau)}{1 + (\omega \tau)^2} \cos (\omega t) \right)
\end{equation}
\begin{equation}
    N_1 = \sigma_{11} - \sigma_{22} = \gamma_0^2 \left( \frac{G (\omega \tau)^2}{1 + (\omega \tau)^2} + \frac{G (\omega \tau)^2 (1 - 2 (\omega \tau)^2)}{(1 + (\omega \tau)^2)(1 + 4(\omega \tau)^2)} \cos (2 \omega t) + \frac{3 G (\omega \tau)^3}{(1 + (\omega \tau)^2)(1 + 4(\omega \tau)^2)} \sin (2 \omega t) \right) \\
\end{equation}
\end{widetext}
where the Deborah number $De = \omega \tau$ naturally arises, and it is clear that the results are time-strain separable.

All material functions of the UCM model, which are the coefficients of the individual trigonometric terms in the equation above, are functions only of the frequency $\omega$:
\begin{subequations}
\begin{align}
\begin{split}
    G'(\omega) &= \frac{G (\omega \tau)^2}{1 + (\omega \tau)^2} 
\end{split} \\
\begin{split}
    G''(\omega) &= \frac{G (\omega \tau)}{1 + (\omega \tau)^2} 
\end{split}\\
\begin{split} \label{n1d}
    n_1^d (\omega) &= \frac{G (\omega \tau)^2}{1 + (\omega \tau)^2} 
\end{split}\\
\begin{split} \label{n1prime}
    n_1' (\omega) &= \frac{G (\omega \tau)^2 (1 - 2 (\omega \tau)^2)}{(1 + (\omega \tau)^2)(1 + 4(\omega \tau)^2)} 
\end{split} \\
\begin{split}
    n_1'' (\omega) &= \frac{3 G (\omega \tau)^3}{(1 + (\omega \tau)^2)(1 + 4(\omega \tau)^2)}
\end{split}
\end{align}
\end{subequations}

Notably, all of these coefficients can be expressed in terms of the two material functions determined from linear viscoelasticity \cite{ferry_viscoelastic_1980}:
\begin{equation}
    \sigma_{12} = \gamma_0 \left(G' (\omega) \sin (\omega t) + G'' (\omega) \cos (\omega t) \right)
\end{equation}
%\begin{equation}
%    N_1 = \gamma_0^2 \biggl( \underbrace{G' (\omega)}_{n_1^d (\omega)} + \Bigl( \underbrace{-G'(\omega) + \tfrac{1}{2} G' (2\omega)}_{n_1'(\omega)} \Bigl) \cos (2 \omega t) + \Bigl( \underbrace{G''(\omega) - \tfrac{1}{2} G''(2\omega)}_{n_1''(\omega)} \Bigl) \sin (2 \omega t) \biggl)
%\end{equation}
\begin{eqnarray}
    N_1 &=& \gamma_0^2 \biggl( \underbrace{G' (\omega)}_{n_1^d (\omega)} + \Bigl( \underbrace{-G'(\omega) + \frac{1}{2} G' (2\omega)}_{n_1'(\omega)} \Bigl) \cos (2 \omega t) \nonumber\\
    & &+ \Bigl( \underbrace{G''(\omega) - \frac{1}{2} G''(2\omega)}_{n_1''(\omega)} \Bigl) \sin (2 \omega t) \biggl)
\end{eqnarray}

Even when the strain rate is low (in oscillatory shear, this will be true for a sufficiently low $\gamma_0$), normal stress differences are present, but they are too small to be detected because of the asymptotic quadratic dependence on $\gamma_0$. It is also notable that for the quasilinear UCM model, the material functions for $N_1$ are directly related to the differences between $G'$ and $G''$ at two different frequencies. The presence of normal stress differences is often thought of as a purely elastic phenomenon. This could be due to the fact that normal forces are often measured in simple shear only when the applied strain rate is sufficiently high. The simple analysis above shows that this is not the full picture - normal stress differences are \textit{nonlinear} viscoelastic phenomena that have both elastic and also viscous contributions that are difficult to separate \cite{yu_general_2009}. At second order, the $N_1$ coefficients are directly related to the elastic and viscous coefficients obtained from the shear stress (i.e. to $G^* (\omega) = G' (\omega) + iG'' (\omega)$), but the normal stress differences do not contribute to energy dissipation over one cycle of deformation. 

In the new Fourier-Tschebyshev notation that we propose, the LAOS material functions predicted by the UCM model are only a function of the applied angular frequency and not the strain amplitude:
\begin{subequations} \label{notation}
\begin{align}
\begin{split}
    n_1^d &= \frac{G (\omega \tau)^2}{1 + (\omega \tau)^2} 
\end{split} \\
\begin{split}
    |n_{1,2}^*| &= \sqrt{(n_{1,2}')^2 + (n_{1,2}'')^2} = \frac{G (\omega \tau)^2}{\sqrt{(1+(\omega \tau)^2)(1+4(\omega \tau)^2)}} 
\end{split} \\
\begin{split}
    \Phi_2 &= \tan^{-1} \left(\frac{n_{1,2}''}{n_{1,2}'} \right) = \tan^{-1} \left(\frac{3(\omega \tau)}{1 - 2(\omega \tau)^2} \right)
\end{split}
\end{align}
\end{subequations}

It is interesting to note that $\tan \Phi_2$ can change sign at a critical Deborah number $De_c = \tau \omega_c = \sqrt{1/2}$, which is the point at which $\Phi_2 = \pi/2$. This is primarily because the value of $n_1' (\omega)$ changes sign at that critical frequency (see equation \ref{n1prime}). 

\begin{figure*}
\centering
\includegraphics[width=\textwidth]{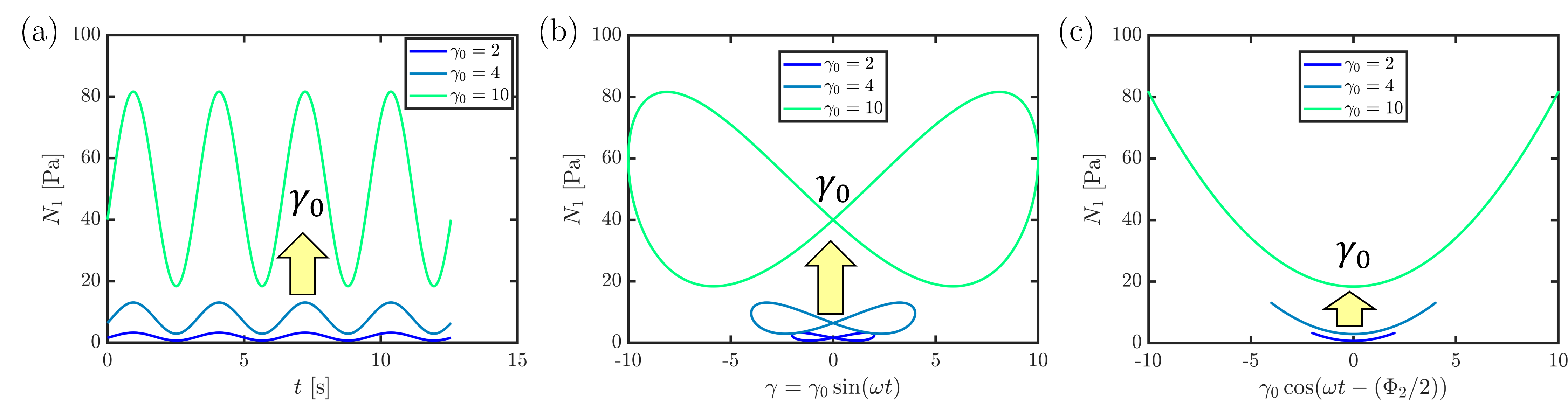}
\caption{The evolution of $N_1 (t)$ calculated using the Upper Convected Maxwell model (here $G = 1$ Pa, $\tau = 1$ s, $\omega = 1$ rad/s) for different strain amplitudes, presented as (a) time series data, (b) Lissajous figures of the first normal stress difference vs. strain, and (c) single-valued curves with an abscissa $z = \cos(\omega t - \Phi_2/2)$ that incorporates the phase angle $\Phi_2$. In this specific 2D projection of the (naturally 3D) Lissajous orbits, curves of $N_1(\omega,\gamma_0)$ overlap and enclose no area.} 
\label{fig:UCM_amplitude}
\end{figure*}

Lissajous curves often offer a more visually accessible interpretation of stress-strain relationships rather than a plot of time-series data. In contrast to the largely elliptical shapes of shear stress Lissajous curves, the $N_1$ orbits $\{\gamma(t); N_1(t)\}$ are `butterfly-shaped' curves with a non-zero mean corresponding to the DC term $n_1^d (\omega)$ (see equation \ref{n1d}). It is important to note that the UCM model is a quasilinear model that only generates the zeroth and second order harmonics of the general expansion given in equations \ref{definition1} and \ref{definition}. An increase in the strain amplitude $\gamma_0$ only leads to the Lissajous curve growing in size, with no change to its shape (\textbf{Figure \ref{fig:UCM_amplitude}}). Plotting the $N_1$ data with respect to a new abscissa $z = \cos(\omega t - \Phi_2/2)$ that incorporates the phase angle $\Phi_2$ leads to a single-valued function $T_2(z)$ which is quadratic in $z$. This gives a geometrical meaning to the normal stress phase angle $\Phi_2$; it is the angle of phase delay that results in the $N_1 (t; \omega, \gamma_0)$ signal becoming a single-valued function.

\begin{figure*}
\centering
\includegraphics[width=\textwidth]{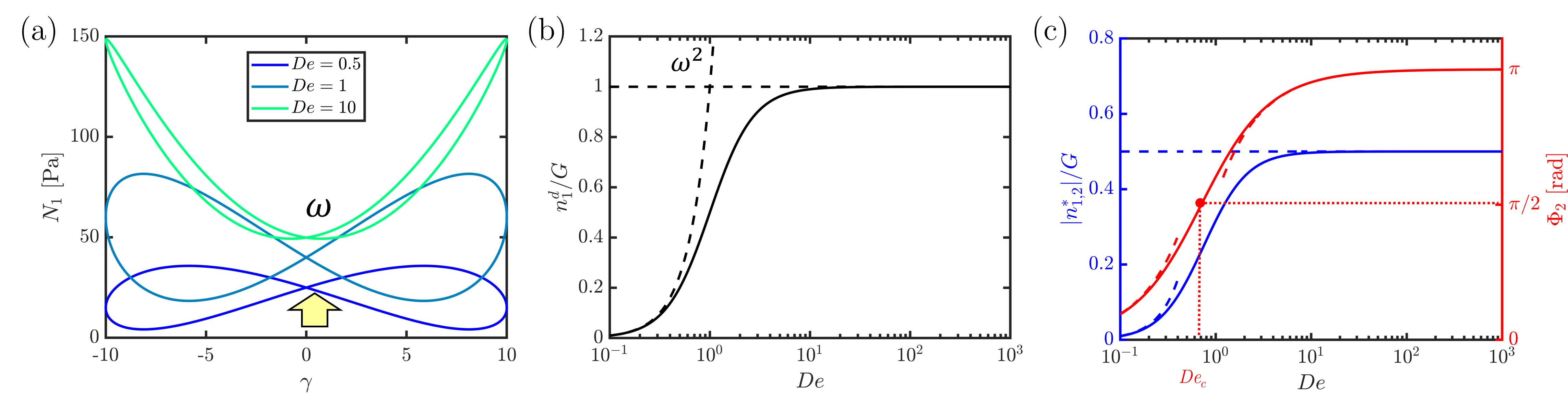}
\caption{(a) The evolution of $N_1$ calculated using the Upper Convected Maxwell model ($G = 1$ Pa, $\tau = 1$ s, $\gamma_0 = 10$) for different angular frequencies, presented as stress-strain Lissajous curves. (b) and (c) Analytical solutions for the zeroth and second harmonics of the UCM model as well as asymptotic values given in the text. Note the change in sign for $\tan \Phi_2$ at a critical Deborah number $De_c = \omega_c \tau = \sqrt{1/2}$. The asymptotic expressions for $n_1^d$, $|n_{1,2}^*|$ and $\Phi_2$ at small and large $De$ (equations \ref{UCMsmallDe_n1d} to \ref{UCMasymptotic_end}) are marked as dashed lines.} 
\label{fig:UCM_angle}
\end{figure*}

On the other hand, an increase in the angular frequency $\omega$ of the applied deformation leads to a change in the material functions, modifying both the size and shape of the corresponding Lissajous curves as shown in \textbf{Figure \ref{fig:UCM_angle}}. Because analytical expressions are available for the UCM model, we are able to determine how the material functions vary with $De$. In the limit of small $De$, the UCM model approximates a second-order fluid, in which the coefficients $n_1^d$ and $|n_{1,2}^*|$ both vary as the square of the angular frequency $\omega$:
\begin{equation} \label{UCMsmallDe_n1d}
    \lim_{De \ll 1} n_1^d (\omega) = G (\omega \tau)^2
\end{equation}
\begin{equation} \label{UCMsmallDe_n12}
    \lim_{De \ll 1} |n_{1,2}^*| (\omega) = G (\omega \tau)^2
\end{equation}
\begin{equation}
    \lim_{De \ll 1} \Phi_2 (\omega) = 3 \omega \tau
\end{equation}
In the limit of zero $De$, all coefficients vanish asymptotically, in agreement with the fact that a Newtonian fluid does not exhibit normal stress differences. 

Conversely, in the limit of large $De$, 
\begin{equation}\label{UCMlargeDe_n1d}
    \lim_{De \gg 1} n_1^d (\omega) = G \left(1 - \frac{1}{(\omega \tau)^2} + ... \right)
\end{equation}
\begin{equation} \label{UCMlargeDe_n12}
    \lim_{De \gg 1} |n_{1,2}^*| (\omega) = \frac{G}{2} \left(1 - \frac{5}{8 (\omega \tau)^2} + ... \right)
\end{equation}
\begin{equation} \label{UCMasymptotic_end}
    \lim_{De \gg 1} \Phi_2 (\omega) = \pi - \frac{3}{2 \omega \tau} + ...
\end{equation}
In the limit of moderate strains, but large $De$ (i.e. in the weakly nonlinear elastic limit), the $N_1$ coefficients thus become constants that are independent of $De$ (see \textbf{Figure \ref{fig:UCM_angle}(c)}), and $N_1 \simeq G \gamma_0^2 (1 - \tfrac{1}{2} \cos (2\omega t))$. Normal stress differences that arise in large amplitude oscillatory shearing are similar to the normal stresses that lead to the axial extension of a solid when subjected to torsion (also known as the Poynting effect) observed in the field of finite elasticity in nonlinear solid mechanics \cite{anand_continuum_2020}. Normal stress differences were also reported in a recent LAOS study on meat products \cite{gimenez-ribes_effect_2024}, which are soft viscoelastic solids. The authors defined a material function known as the "Poynting modulus" to quantify the normal stresses. For the simplest nonlinear elastic solid model, a neo-Hookean solid, $N_1$ is given by:
\begin{eqnarray}
    N_1 &\equiv& G \gamma^2 = G \gamma_0^2 \sin^2 (\omega t) = \frac{G}{2} \gamma_0^2 (1 - \cos (2\omega t)) \nonumber \\
    &\equiv& G \gamma_0^2 \left(1 - \tfrac{1}{2} \cos(2\omega t) \right) - \tfrac{1}{2} G \gamma_0^2
\end{eqnarray}
This is satisfyingly consistent with the limit of the UCM model at large \textit{De}, except for the constant offset term. A difference in the reference configuration of solids as compared to liquids explains this discrepancy. 

In light of the results above, the reason behind the switch in notation as compared to Nam \cite{nam_prediction_2008,nam_first_2010} becomes apparent: the new notation we propose in equation \ref{propose} is now consistent with the convention that oscillatory material functions with a single prime symbolize elastic characteristics (here multiplying the $\cos(2\omega t)$ term) and those with a double prime symbolize viscous characteristics. For $De \rightarrow \infty$, we thus have $n_1' = -G/2$ and $n_1'' = 0$. 

It is often the case that a discrete Fourier transform (DFT) is applied to the strain and stress signals. The DFT of a function $x(t_m)$ is defined as:
\begin{equation}
    \tilde{x}_{DFT} (\omega_k) = \sum_{m=0}^{M-1} x(t_m) e^{-i\omega_k t_m}
\end{equation}
where $x(t_m)$ is the input signal amplitude at each time $t_m$, $M$ is the total number of points in the discrete time signal, $dt$ is the time sampling interval, and $t_m = m\cdot dt$ is the time corresponding to the $m$th sampling instant.

As we have shown, at low strain amplitudes, the zeroth harmonic material functions of $N_1$ are related to the shear stress material functions by:
\begin{equation}
    \omega^2 \Psi_1^d = n_1^d = G' = \omega \eta''
\end{equation}
When a discrete Fourier transform (DFT) is taken of the measured time series $N_1(t)$, the complex first normal stress coefficients are defined as:
\begin{equation}
    \Psi_1^* = \frac{\tilde{N_1}}{(\tilde{\dot{\gamma}})^2} M = \Psi_1' - i\Psi_1''
\end{equation}
\begin{equation}
    n_1^* = \frac{\tilde{N_1}}{(\tilde{\gamma})^2} M = - n_1' + i n_1''
\end{equation}

where the $\sim$ notation above a symbol indicates a DFT-transformed variable. These expressions are related to each other through the following relation: $n_1^* = (i\omega)^2 \Psi_1^*$. Note that unlike the case of the shear stress coefficients $G^*$ and $\eta^*$, the coefficients here depend on the number of samples $M$ in the discrete transformed signal, because of the quadratic dependence on the strain signal. In the Fourier domain, the following two relations are also valid at low $\gamma_0$:
\begin{equation} \label{commonality_start}
    i\omega \Psi_1^* = \eta^* (\omega) - \eta^* (2\omega)
\end{equation}
\begin{equation}
    n_1^* = G^* (\omega) - \tfrac{1}{2} G^* (2\omega)
\end{equation}

When separated into real and imaginary contributions, this yields the following relations:
\begin{equation}
    \omega \Psi_1' = - \eta''(\omega) + \eta''(2\omega)
\end{equation}
\begin{equation}
    \omega \Psi_1'' = \eta'(\omega) - \eta'(2\omega)
\end{equation}
\begin{equation}
    n_1' = -G'(\omega) + \tfrac{1}{2} G'(2\omega)
\end{equation}
\begin{equation} \label{commonality_end}
    n_1'' = G''(\omega) - \tfrac{1}{2} G''(2\omega)
\end{equation}

As seen above, there are clear relations between the linear viscoelastic coefficients and the $N_1$ coefficients at second order. It must be noted that this is not unique to the UCM model, but is valid even if the upper convected derivative is replaced by the general Gordon-Schowalter derivative $\accentset{\square}{\pmb{\sigma}}_p \equiv \xi \accentset{\triangledown}{\pmb{\sigma}}_p + (1-\xi) \accentset{\triangle}{\pmb{\sigma}}_p$, which is a weighted sum of the upper and lower convected derivatives. The upper convected derivative is a special case in which $\xi = 1$. Another special case is the lower convected derivative $\accentset{\triangle}{\pmb{\sigma}}_p \equiv \frac{D \pmb{\sigma}_p}{D t} + (\nabla \pmb{v})^T \cdot \pmb{\sigma}_p + \pmb{\sigma}_p \cdot \nabla \pmb{v}$, in which $\xi = 0$, producing the Lower Convected Maxwell model, which also has the exact same expression for $N_1$. Other special cases, such as the Co-Rotational Maxwell Model (where $\xi = 1/2$), produce an infinite series of even harmonics for $N_1$, but the zeroth and second harmonics at low $\gamma_0$ are the same as were calculated by Giacomin and coworkers \cite{giacomin_large-amplitude_2011}. The general Gordon-Schowalter derivative is incorporated into a number of popular models, such as the Johnson-Segalman model \cite{johnson_model_1977} and the original Phan-Thien---Tanner model \cite{phan-thien_new_1977}. This commonality in the response is because the identities in equations \ref{commonality_start} to \ref{commonality_end} arise from second-order fluid theory and the $\tfrac{D}{Dt}(\cdot)$ operator rather than the frame-invariant contributions.

These identities provide assurance that the $N_1$ Fourier-Tschebyshev framework introduced here will be valid for a broad class of fluids. Additionally, the fact that there are asymptotic relations between the material functions for the shear stress and $N_1$ provides a self-consistency check that fluids must obey at low strain amplitudes, i.e. in the quasilinear regime, which is helpful for checking the calibration of rheometric instrumentation used to collect experimental data. To reiterate, the self-consistency checks are:
\begin{equation} \label{consistency1}
    \lim_{\gamma_0 \to 0} n_1^d (\omega) = \lim_{\gamma_0 \to 0} G_1'(\omega) = G'(\omega)
\end{equation}
\begin{equation} \label{consistency2}
    \lim_{\gamma_0 \to 0} n_1'(\omega) = \lim_{\gamma_0 \to 0} \left(-G_1'(\omega) + \frac{1}{2} G_1' (2\omega) \right)
\end{equation}
\begin{equation} \label{consistency3}
    \lim_{\gamma_0 \to 0} n_1'' (\omega) = \lim_{\gamma_0 \to 0} \left( G_1'' (\omega) - \frac{1}{2} G_1'' (2\omega) \right)
\end{equation}

\section{Gaborheometry applied to $N_1$ in the Giesekus model}

The great advantage of large amplitude oscillatory shear experiments is the ability to independently vary an observational timescale (parameterized by $\omega$) and the nonlinearity (parameterized by $\gamma_0$) separately. Obtaining the values of each harmonic intensity and phase angle for each combination of angular frequency and strain amplitude $(\omega, \gamma_0)$, or in dimensionless terms, $(De, \gamma_0)$ is laborious. A technique known as Gaborheometry strain sweeps (GaborSS) was recently developed \cite{rathinaraj_gaborheometry_2023}, in which a ramped oscillatory deformation of the general form $\gamma(t) = \gamma_0 (t) \sin(\omega t)$ with a time-varying strain amplitude $\gamma_0 (t) = \gamma_i + rt$, where $\gamma_i$ is the initial strain amplitude and $r$ is the linear ramp rate, is applied to a sample. The resultant stress signal is then decomposed and localized in time (and thus in strain amplitude) using a Gabor transform, obtaining the evolving material functions for a single frequency, across multiple strain amplitudes, within a single experiment. This significantly shortens the time needed to explore the entire Pipkin space. A brief review of the technique is given below.

The discrete Gabor transform (DGT) of a function $x$, at a time of interest $\tau$ and frequency of interest $\omega$, is:
\begin{equation}
    \tilde{x}_{DGT} (\tau,\omega_k) = \sum_{m=0}^{M-1} x(t_m) g(t_m-\tau) e^{-i\omega_k t_m}
\end{equation}
where $x(t_m)$ is the input signal amplitude at each time $t_m$, $M$ is the total number of points in the discrete time signal, $dt$ is the time sampling interval, and $t_m = m\cdot dt$ is the time corresponding to the $m$th sampling instant.
The Gaussian window function localizing the material response in time is:
\begin{equation}
    g(t_m-\tau) = A_w e^{-(t_m-\tau)^2 / 2a^2}
\end{equation}
The normalization factor is:
\begin{equation}
    A_w = \frac{M}{\sum_{m=0}^{M-1} g(t_m)}
\end{equation}
The optimal window length for the frequency of interest $\omega_i$ was a principal result of the original study by Rathinaraj and McKinley \cite{rathinaraj_gaborheometry_2023}:
\begin{equation}
    a =\frac{2.63}{\omega_i}
\end{equation}

\begin{figure*}
\centering
\includegraphics[width=0.8\textwidth]{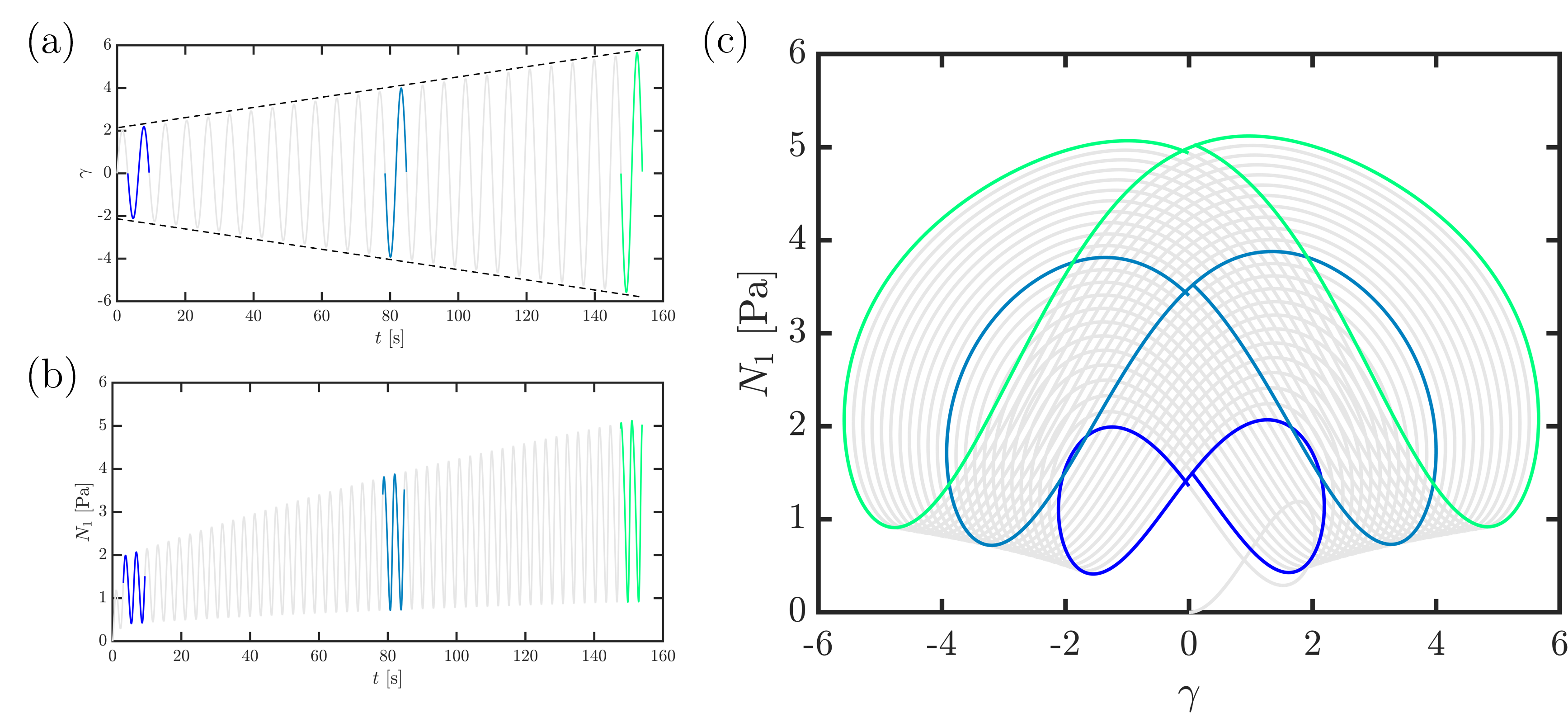}
\caption{The input to the Giesekus model ($G = 1$ Pa, $\tau = 1$ s, $\omega = 1$ rad/s, $\alpha = 0.4$) is (a) an amplitude-modulated strain sweep that is linearly increasing in strain amplitude, producing (b) an oscillatory normal stress difference signal that oscillates about a non-zero mean at a frequency $2\omega$. (c) The resultant Lissajous curves of $N_1$ grow in size with time. These signals can be analyzed using the windowed Gabor transform to obtain material functions for $N_1$ in LAOS at each strain amplitude applied. Three representative discrete single cycles at $t_i = 9.425, 84.82, 153.9$ s (corresponding to strain amplitudes $\gamma_0 (t_i) = 2.226, 4.035, 5.694$) are highlighted in green and blue to show the progression in the size and shape of the curves with time.} 
\label{fig:gabor_explanation}
\end{figure*}

Because this experimental protocol is not time-translation invariant, it is important to monitor the rate of change in the imposed oscillatory signal \cite{rathinaraj_gaborheometry_2023}. The mutation number is a ratio of the characteristic time constant for the change of material (in this case, the reciprocal of the ramp rate of the strain amplitude function) to the relevant experimental time (the period of one cycle of oscillation)\cite{mours_time-resolved_1994}. For a Gaborheometry strain sweep, it is thus appropriately defined as $Mu = \frac{2\pi}{\omega} \frac{r}{\gamma_0}$. Crucially, the mutation number $Mu$ must be kept small (Rathinaraj and McKinley \cite{rathinaraj_gaborheometry_2023} suggested $Mu< 0.1$) in order for the computed results of $G^* (\omega, t)$ to be accurate. The Gaborheometry strain sweep method has previously been applied to obtain the strain-dependent elastic and viscous Fourier-Tschebyshev coefficients for the shear stress signal in LAOS. We show here that the technique can also be adapted to obtain the newly defined $N_1$ coefficients in LAOS, first for a well-defined model that exhibits multiple harmonic contributions (i.e. a fully non-linear model such as the Giesekus model), and then to real experimental data measured in a weakly elastic silicone calibration fluid and then a strongly elastic thermoplastic polyurethane (TPU) melt. As shown in \textbf{Figure \ref{fig:gabor_explanation}}, the ramped strain signal produces $N_1$ Lissajous curves which grow in size, in the magnitude of their DC offset and nonlinearity. This single strain sweep can then be systematically analyzed with the Gabor window function to obtain the relevant material functions at each instant in time (or equivalently, strain amplitude).

The Giesekus model \cite{giesekus_simple_1982} is commonly used to model the nonlinear rheological properties of a wide range of polymer melts. Assuming no solvent contribution, the Giesekus model for the polymeric stress is given by the following expression: 
\begin{equation}
    \tau \accentset{\triangledown}{\pmb{\sigma}}_p + \pmb{\sigma}_p + \frac{\alpha}{G} \pmb{\sigma}_p \cdot \pmb{\sigma}_p = G \tau \dot{\pmb{\gamma}}
\end{equation}
From the microstructural standpoint of the bead-spring model of a polymer chain, the parameter $\alpha$ controls the degree of anisotropic drag experienced by the beads of the dumbbell. Setting $\alpha = 0$ recovers the UCM model.

In oscillatory shear, the following set of coupled nonlinear differential equations is obtained:
\begin{subequations} \label{giesekus33}
\begin{align}
\begin{split}
    \frac{d\sigma_{11}}{dt} + \frac{\sigma_{11}}{\tau} + (\sigma_{11}^2 + \sigma_{12}^2) \frac{\alpha}{\tau G} - 2\dot{\gamma} \sigma_{12} &= 0 
\end{split} \\
\begin{split}
    \frac{d \sigma_{22}}{dt} + \frac{\sigma_{22}}{\tau} + (\sigma_{12}^2 + \sigma_{22}^2) \frac{\alpha}{\tau G} &= 0
\end{split} \\
\begin{split}
    \frac{d \sigma_{12}}{dt} + \frac{\sigma_{12}}{\tau} + (\sigma_{11} + \sigma_{22}) \sigma_{12} \frac{\alpha}{\tau G} - \dot{\gamma} \sigma_{22} - G \dot{\gamma} &=0
\end{split} \\
\begin{split}
    \frac{d \sigma_{33}}{dt} + \frac{\sigma_{33}}{\tau} + \sigma_{33}^2 \frac{\alpha}{\tau G} &= 0
\end{split}
\end{align}
\end{subequations}

\begin{figure*}
\centering
\includegraphics[width=\textwidth]{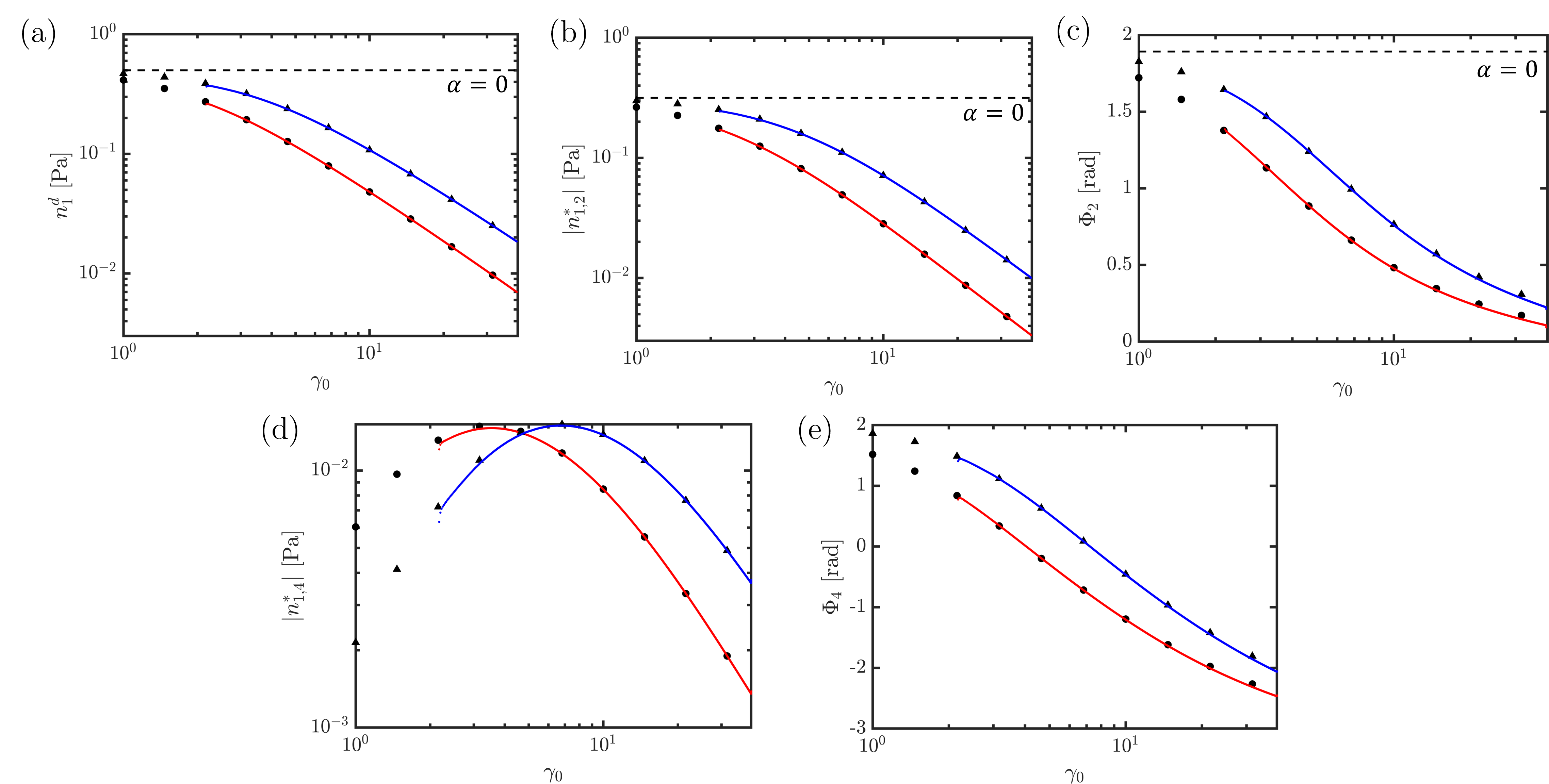}
\caption{Comparison between the discrete strain amplitude simulations (black symbols) and the continuous strain sweep ($\gamma_0(t) = 2 + 0.024t$) simulations of the Giesekus model ($G = 1$ Pa, $\tau = 1$ s, $\omega = 1$ rad/s, with $\alpha = 0.1, 0.4$). The maximum mutation number was $Mu_{\text{max}} = 0.075$ at $t = 0$. The contributions to the zeroth, second, and fourth harmonics of the $N_1$ signal are presented. The predictions of the zeroth and second harmonics from the UCM model (i.e. $\alpha = 0$ in the Giesekus model) are shown as black dashed lines in (a), (b) and (c). Simulations for $\alpha = 0.1$ are shown by the black triangles (discrete imposed strain amplitude) and closely spaced blue dots (GaborSS), while the case when $\alpha = 0.4$ is shown by the black circles (discrete strains) and closely spaced red dots (GaborSS).} 
\label{fig:giesekus_gabor}
\end{figure*}

As is typical, here the index 1 refers to the direction of flow, 2 refers to the direction of the velocity gradient, or shear, and 3 refers to the neutral, or vorticity direction. Assuming the initial conditions correspond to a rest state with $\pmb{\sigma} = \pmb{0}$, equation \ref{giesekus33}(d) indicates that we can trivially set $\sigma_{33} = 0$ at all times. However, the presence of the quadratic stress terms in each of the remaining equations leads to a system of three coupled nonlinear differential equations, which have been solved numerically for a time- and frequency-varying input $\gamma (t) = (\gamma_i + rt) \sin (\omega t)$. The complex time-varying limit cycle shown in Figure \ref{fig:gabor_explanation}(c)) can be analyzed rapidly using the Gabor transform. As we show in \textbf{Figure \ref{fig:giesekus_gabor}}, the Gaborheometry strain sweep technique produces good agreement with a series of discrete strain amplitude simulations (black discrete points) in the Giesekus model, as long as the mutation number $Mu$ is kept small. As the imposed strain amplitude $\gamma_0$ increases, strain softening in the computed material response occurs and the intensities of $n_1^d$ and $|n_{1,2}^*|$ decrease. Meanwhile, the intensity of $|n_{1,4}^*|$ is predicted to pass through a maximum at a critical ($\alpha$-dependent) strain amplitude. A lower value of $\alpha$ leads to a more gradual decrease in the values of $n_1^d$ and $|n_{1,2}^*|$ with $\gamma_0$. This is consistent with the fact that when $\alpha = 0$, the UCM model is recovered, and the values of $n_1^d$ and $|n_{1,2}^*|$ will not vary with strain amplitude, instead producing horizontal lines as shown in \textbf{Figure \ref{fig:giesekus_gabor}(a)} to \textbf{(c)}. 

\begin{figure*}
\centering
\includegraphics[width=\textwidth]{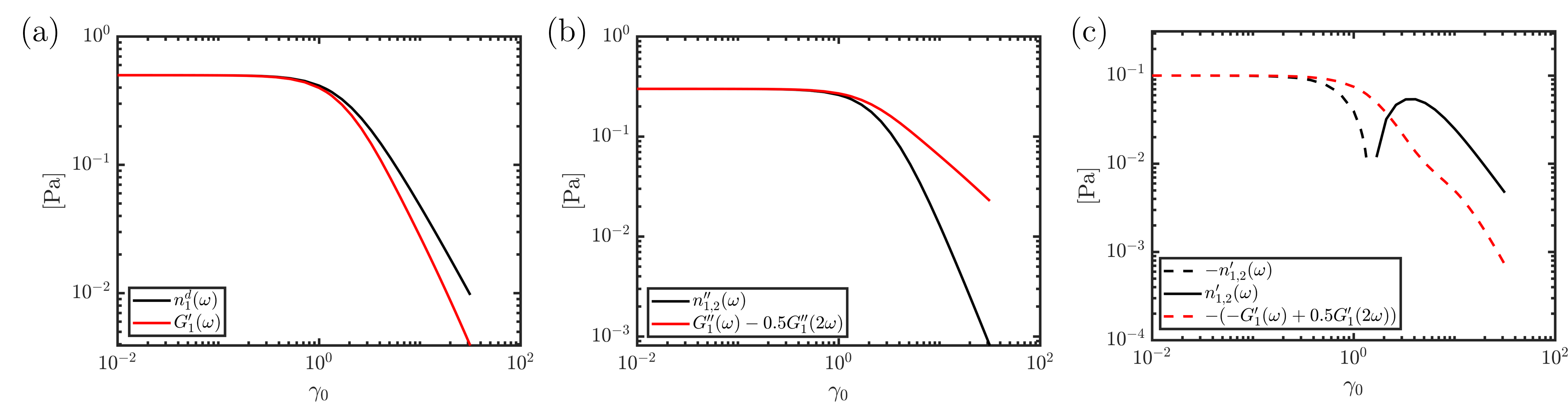}
\caption{The asymptotic low $\gamma_0$ limit for the zeroth and second harmonics of $N_1$ signal predicted by the Giesekus model ($G = 1$ Pa, $\tau = 1$ s, $\omega = 1$ rad/s, $\alpha = 0.4$) match well with the appropriate combinations of the first harmonic measurements of the complex modulus. In each figure, the black line is computed from the ramped oscillatory $N_1$ data while the red line is obtained from the oscillatory shear stress data. Dotted lines in (c) indicate negative values of the coefficient due to a change in sign of $n_{1,2}'(\omega)$ at a critical strain amplitude.} 
\label{fig:giesekus_asymptotic}
\end{figure*}

\begin{figure*}
\centering
\includegraphics[width=\textwidth]{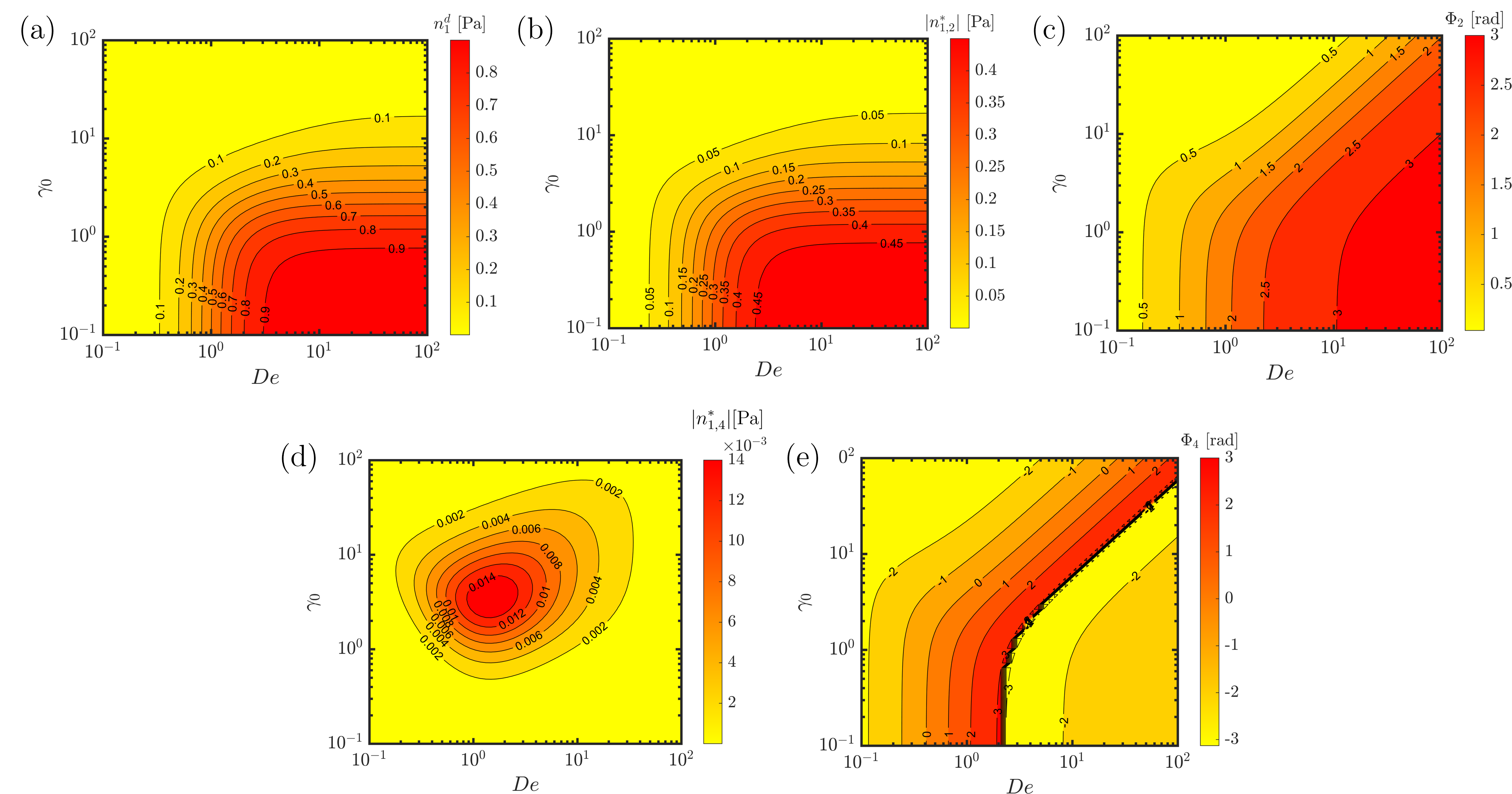}
\caption{Pipkin diagrams showing the evolution in the $N_1$ LAOS coefficients for the Giesekus model ($G=1$ Pa, $\tau = 1$ s, $\omega = 1$ rad/s, $\alpha = 0.4$) as a function of angular frequency ($De = \omega \tau$) and strain amplitude ($\gamma_0$). The magnitude of the DC coefficient $n_1^d$ is shown in (a) and then the strength and phase information for the second and fourth harmonics (using the notation of equation \ref{notation}) are shown in (b), (c) and (d), (e) respectively. At low $\gamma_0$, the Giesekus model reduces to the UCM model, and there is good agreement for the zeroth and second harmonics with the analytical results of the UCM model, which are each a function of angular frequency only.} 
\label{fig:giesekus_contour}
\end{figure*}

In the limit of low $\gamma_0$, we expect the self-consistency checks summarized in Equations \ref{consistency1} to \ref{consistency3} to hold. We see excellent agreement between the oscillatory shear stress and $N_1$ LAOS data in that quasilinear limit (\textbf{Figure \ref{fig:giesekus_asymptotic}}). This is unsurprising because small values of $\gamma_0$ lead to the quadratic stress term in the Giesekus model becoming negligibly small, recovering the limiting case of the quasilinear UCM model. As was shown in the previous section, the analytical solutions of the UCM model satisfy the asymptotic relations directly at all strains. At moderate $\gamma_0$, an analytical solution for the Medium Amplitude Oscillatory Shear (MAOS) expansion up to fourth order in $\gamma_0$ is possible. This was obtained using a combination of manual calculations and with the aid of Mathematica, following the approach taken by Gurnon and Wagner \cite{gurnon_large_2012} for the shear stress. The full expressions up to $\mathcal{O} (\gamma_0^4)$ are given in the Appendix, and the numerical simulations match the analytical solutions very well, fostering confidence in the Gaborheometry process for extracting higher order harmonics.

A useful representation summarizing nonlinear viscoelastic behavior is the Pipkin diagram. This is a two-dimensional graph with the Deborah number on the abscissa and the strain amplitude on the ordinate. Other representations may include the Weissenberg number, which is a dimensionless strain rate $Wi = \gamma_0 \omega \tau$, or the stress amplitude $\sigma_0$ (for controlled-stress experiments) on the $y$-axis \cite{ewoldt_mapping_2017}. Pipkin diagrams created as contour plots of each of the Fourier-Tschebyshev coefficients for $N_1$ in LAOS act as \textit{rheological fingerprints} of a fluid, and complement the Pipkin diagrams obtained for the corresponding coefficients from oscillatory shear stress data. The corresponding Pipkin diagrams for the Giesekus model with $\alpha = 0.4$ are shown in \textbf{Figure \ref{fig:giesekus_contour}}. The computed results show good agreement with the UCM model at low $\gamma_0$, and this is reflected graphically by the contour lines in \textbf{Figure \ref{fig:giesekus_contour} (a)} to \textbf{(c)} becoming vertical (i.e. independent of strain amplitude) for $\gamma_0 \ll 1$. The change in the color intensity of the contour lines at low $\gamma_0$ (i.e. moving horizontally across the Pipkin diagram) for the zeroth and second harmonics directly matches the graphs obtained in \textbf{Figure \ref{fig:UCM_angle}}. Although the maximum value of $|n_{1,4}^*|$ is an order of magnitude less than $n_1^d$, a larger value of $\alpha$ will produce more pronounced nonlinear behavior and shift the contour lines of all diagrams downwards, i.e. the fluid begins exhibiting nonlinear behavior at a lower $\gamma_0$. 

Through these analytical and computational considerations, we have shown that the Gaborheometry technique originally proposed by Rahinaraj and McKinley \cite{rathinaraj_gaborheometry_2023} for measuring the time-localized rheological properties of mutating materials can be adapted to obtain the $N_1$ material functions in LAOS as a function of a time-varying oscillatory strain amplitude $\gamma_0(t)$ using the protocol defined in the preceding section. Furthermore, the self-consistency checks between the oscillatory shear stress and $N_1$ signals were demonstrated to be satisfied using the Giesekus model, and the higher harmonics, which provide a rheological fingerprint of the nonlinear viscoelastic material response, can also be extracted and plotted on a Pipkin diagram.

\section{Experimental data to illustrate the $N_1$ framework}

Several experimental complexities arise when attempting to use large amplitude oscillations to measure $N_1$ in a fluid. We need to perform experiments at low $\gamma_0$ (the SAOS to MAOS regime) to confirm the self-consistency between shear stress and $N_1$ signals of the fluid. Such measurements should ideally be performed with a cone-and-plate geometry such that $N_1$ can be directly measured. However, to obtain a more complete rheological fingerprint of the material, we want to go to higher $\gamma_0$ values (the LAOS regime) such that the intensities of the higher harmonics are above the noise floor of the force transducer of the rheometer. This is often not possible with a cone-and-plate geometry because highly viscoelastic samples will be ejected from the gap, or edge fracture will occur. To circumvent these issues, a wide variety of alternative test geometries have been developed, most notably the cone and partitioned plate (CPP) geometry \cite{costanzo_review_2024}. The CPP geometry helps constrain the fluid within the gap and delays the effect of edge fracture. The downside is that an "apparent" $N_1$ signal, rather than a true $N_1$ signal, is obtained instead \cite{meissner_measuring_1989,schweizer_measurement_2002,costanzo_review_2024}. In view of this, we will first demonstrate the self-consistency test with a PDMS silicone polymer, and then subsequently show that nonlinear analysis of higher harmonics is possible with a molten thermoplastic polyurethane (TPU). PDMS is a well-behaved room temperature viscoelastic fluid that is commonly used for rheometric calibration purposes, while TPU is a rheologically complex copolymer widely used in many industries \cite{yilgor_critical_2015} and typically processed in the molten state at high temperatures. The microstructure of TPU comprises alternating hard and soft blocks of different polymer chemistries, giving rise to interesting rheological properties at different temperatures. However, its rheology is not as well-characterized as other industrial polymers such as polystyrene or polyethylene \cite{lu_comparing_2004,yoon_effect_2000,silva_rheological_2015}. The highly nonlinear properties of TPUs are of industrial interest for the production of rubber substitutes, shoes, electronic and medical devices, and other consumer products.

\subsection{Materials}

The PDMS fluid was obtained from TA Instruments as a calibration fluid, with a zero shear viscosity of $\eta_0 = 1005$ Pa.s at $T=25^o$C. The TPU used in the experiments is Estane \textsuperscript{\textregistered} TPU 58206, an aromatic polyester-based thermoplastic polyurethane (TPU) grade provided by Lubrizol. Oscillatory shear experiments were performed on an ARES-G2 rheometer. A 25 mm 0.1 rad cone and plate geometry was used for the PDMS measurements. To minimize artifacts associated with edge fracture, a cone and partitioned plate (CPP) geometry from TA Instruments was used for the TPU measurements, with an outer plate diameter of 25 mm and an inner plate diameter ($2 R_{stem}$) of 10 mm. 

\subsection{Small Amplitude Oscillatory Shear (SAOS)}

\begin{figure*}
\centering
\includegraphics[width=0.8\textwidth]{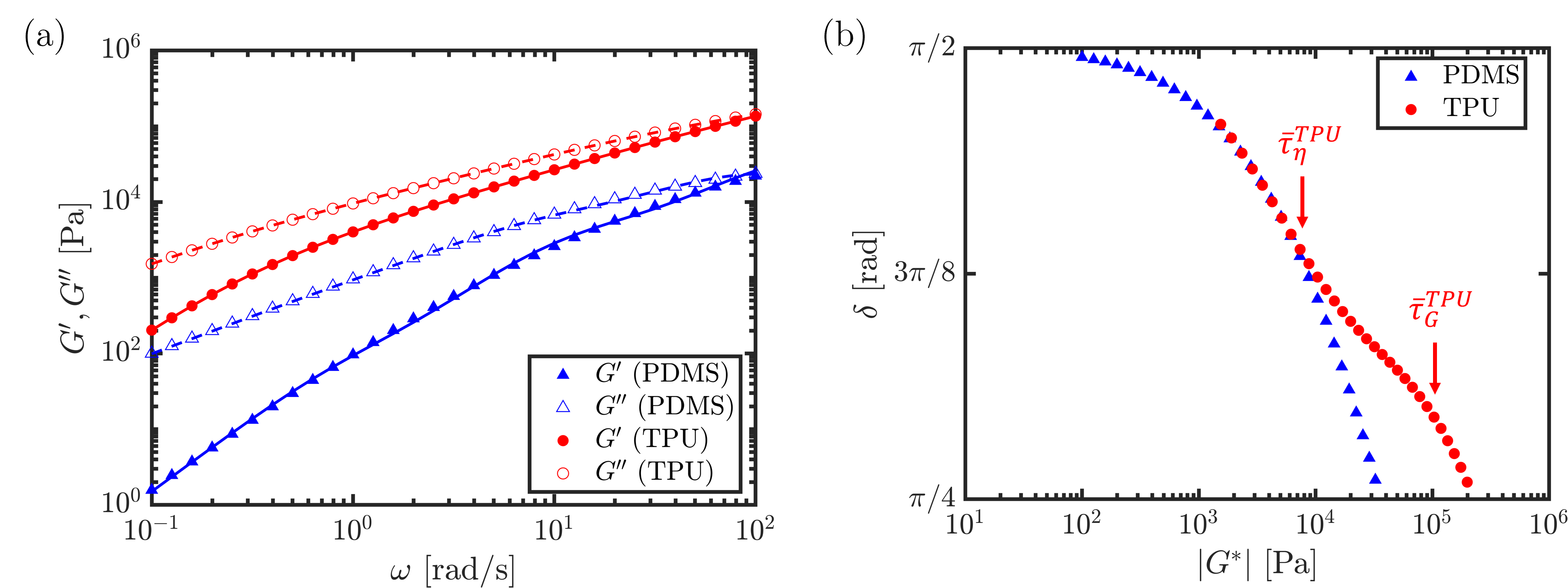}
\caption{Small amplitude oscillatory shear experiments were performed to obtain the characteristic relaxation time of the PDMS silicone fluid at 25$^o$C and the TPU at 190$^o$C. (a) The frequency sweep data for PDMS was fitted to a 5-mode generalized discrete Maxwell spectrum, while the data for the TPU was fitted to a 6-mode generalized discrete Maxwell spectrum. (b) The same data is presented as a van Gurp-Palmen plot, revealing that TPU has two distinct sets of relaxation processes and these can be associated with the two time scales $\bar{\tau}_G^{TPU}$ and $\bar{\tau}_\eta^{TPU}$ respectively.} 
\label{fig:SAOS}
\end{figure*}

Small Amplitude Oscillatory Shear tests were first performed to obtain an estimate of the characteristic relaxation time for each test fluid. This provides guidance on the range of useful angular frequency values that will subsequently be used for LAOS tests. After determining the linear viscoelastic limit for each material, a frequency sweep was performed at 10 \% strain (i.e. $\gamma_0 = 0.1)$ and 25$^o$C for the PDMS, and at 5\% strain (i.e. $\gamma_0 = 0.05$) at 190$^o$C for the molten TPU. We show the resulting SAOS data for both materials in \textbf{Figure \ref{fig:SAOS}}.

Both the PDMS fluid and the TPU produce $G'$ and $G''$ curves that are remarkably similar in shape, and with a similar crossover frequency ($\omega_c \approx 100$ rad/s) as well. The data for both fluids were fitted to a discrete Generalized Maxwell Model using the IRIS software \cite{baumgaertel_determination_1989,winter_cyber_2006}. This produced parsimonious discrete relaxation spectra consisting of five Maxwell modes for the PDMS fluid and six Maxwell modes for the TPU (\textbf{Table \ref{tab:table}}). It is not surprising that the TPU requires more discrete Maxwell modes to produce a good fit to the linear viscoelastic data since it has a broader and flatter material response in SAOS. There are several possible measures for computing a "characteristic" relaxation time from a discrete relaxation spectrum. The modulus-weighted relaxation time $\bar{\tau}_G$ is also a good estimate of the inverse of the crossover frequency, and can be found from the expression:
\begin{equation}
    \bar{\tau}_G = \frac{\sum_{i=1}^N G_i \tau_i}{\sum_{i=1}^N G_i}
\end{equation}
This is computed to be $\bar{\tau}_G = 0.0130$ s for the PDMS fluid and $\bar{\tau}_G = 0.0324$ s for the TPU. As they are so similar, a higher moment of the distribution is required to distinguish the (visually evident) distinct levels of elasticity in the two samples. The more rheologically relevant average relaxation time, which is also the characteristic relaxation time reported from the IRIS software, is the viscosity-weighted relaxation time $\bar{\tau}_\eta$: 
\begin{equation}
    \bar{\tau}_\eta = \frac{\sum_{i=1}^N G_i \tau_i^2}{\sum_{i=1}^N G_i \tau_i}
\end{equation}
This is evaluated to be $\bar{\tau}_\eta = 0.204$ s for the PDMS fluid and $\bar{\tau}_\eta = 1.48$ s for the TPU melt at 190$^o$C.

\begin{table*}
\caption{\label{tab:table} Fits of the experimental data to a multi-mode generalized discrete Maxwell spectrum using the IRIS software for the PDMS fluid (with a zero-shear viscosity at $T = 25^o$C of $\eta_0^{PDMS} = 1005$ Pa.s) and TPU at 190$^o$C (with a zero-shear viscosity of $\eta_0^{TPU} = 15990$ Pa.s)}
\begin{ruledtabular}
\begin{tabular}{ccccc}
Mode $i$ & \multicolumn{2}{c}{PDMS} & \multicolumn{2}{c}{TPU} \\
& Modulus $G_i$ [Pa] & Relaxation Time $\tau_i$ [s] & Modulus $G_i$ [Pa] & Relaxation Time $\tau_i$ [s] \\
\hline
1 & 5.110 & 4.607 & 528.7 & 5.350 \\
2 & 432.9 & 0.3836 & 3002 & 1.529 \\
3 & 3723 & 0.08326 & 8804 & 0.3910 \\
4 & 14890 & 0.02070 & 24300 & 0.09287 \\
5 & 58370 & 0.003367 & 74300 & 0.02157 \\
6 & -& -& 382100 & 0.003335 \\
\hline
$\bar{\tau}_G$ [s]& \multicolumn{2}{c}{0.0130} & \multicolumn{2}{c}{0.0324} \\
$\bar{\tau}_\eta$ [s]& \multicolumn{2}{c}{0.204} & \multicolumn{2}{c}{1.48} \\
$\mathcal{D}$ & \multicolumn{2}{c}{15.7} &\multicolumn{2}{c}{45.7} \\
\end{tabular}
\end{ruledtabular}
\end{table*}

The polydispersity index in terms of the spectrum of relaxation times is a ratio of those two time scales:
\begin{equation}
    \mathcal{D} = \frac{\bar{\tau}_\eta}{\bar{\tau}_G}
\end{equation}
and is evaluated to be $\mathcal{D} = 15.7$ for the PDMS fluid and $\mathcal{D} = 45.7$ for the TPU. This captures the broader spectrum of relaxation times for the TPU melt, which can also be seen in the flatter and broader curves shown in \textbf{Figure \ref{fig:SAOS} (a)}. A useful way of visualizing these differences is the van Gurp-Palmen representation \cite{van_gurp_time-temperature_1998}. This plot is shown in \textbf{Figure \ref{fig:SAOS} (b)}, and the clear `shoulder' evident for the TPU reveals that within the window of frequencies probed, there are at least two distinct relaxation processes for the TPU, each characterized by an associated time scale $\bar{\tau}_G$ and $\bar{\tau}_\eta$. On the other hand, there is only one identifiable relaxation time scale for the PDMS. Clearly, the more complex microstructure of the TPU gives rise to more molecular mechanisms of stress relaxation compared to PDMS, thus producing a more complex relaxation spectrum. 

\subsection{Self-Consistency Checks for PDMS}

\begin{figure}
\centering
\includegraphics[width=0.4\textwidth]{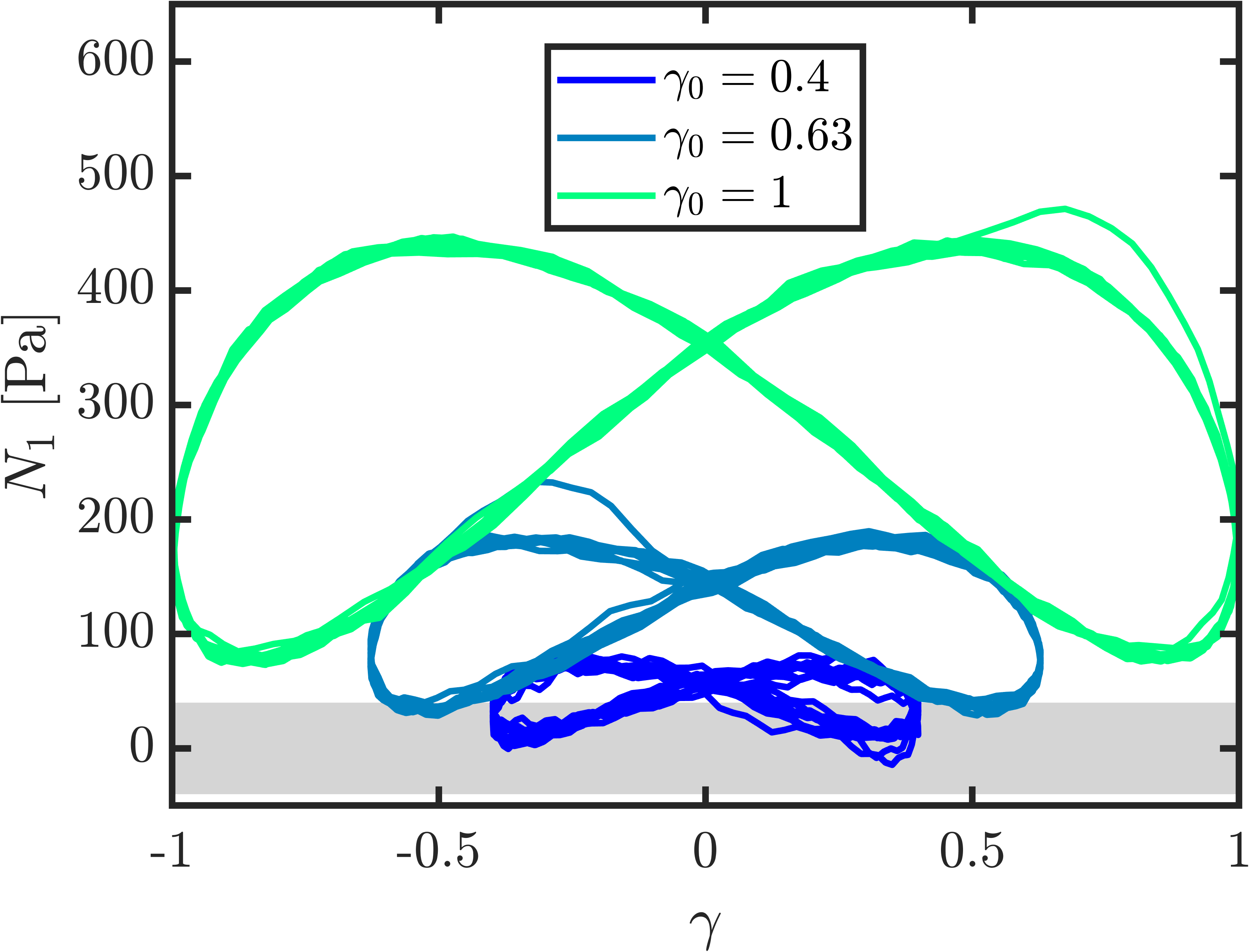}
\caption{Lissajous curves of $N_1$ for the PDMS calibration fluid at room temperature for various $\gamma_0$, at an angular frequency $\omega = 2$ rad/s ($De = \bar{\tau}_\eta \omega = 0.408$). As expected, the curves take the shape of butterfly-shaped curves with a non-zero mean, although the data becomes noisier at lower $\gamma_0$ as they approach the resolution of the normal force transducer. The gray box indicates regions where the measured values of $N_1(t)$ are less than $N_{1,min}$. Larger values of $\gamma_0$ correspond to higher Weissenberg numbers ($Wi = De \gamma_0$) and lead to larger Lissajous curves.} 
\label{fig:PDMS_lissajous}
\end{figure}

Silicone fluids are the prototypical "second-order fluid", which exhibit modest, but measurable normal stress differences and can be considered "weakly elastic". As we show in the Lissajous curves in \textbf{Figure \ref{fig:PDMS_lissajous}}, the magnitudes of the first normal stress difference in LAOS are small, in contrast to the TPU data which will be shown in the next subsection. The minimum axial load that the axial force rebalance transducer of the ARES-G2 can measure is $F_{min} = 1$ gram-force, which corresponds to a normal stress difference of $N_{1, min} = 2F_{min} / \pi R^2 \approx 40$ Pa for a 25 mm plate geometry. This sets the minimum measurable $N_1$ signal that can be adequately resolved above the noise floor and is shown as a gray box in \textbf{Figure \ref{fig:PDMS_lissajous}}. Nevertheless, as expected, the Lissajous curves are distinctly symmetric and butterfly-shaped, as predicted from second-order fluid theory.

\begin{figure*}
\centering
\includegraphics[width=\textwidth]{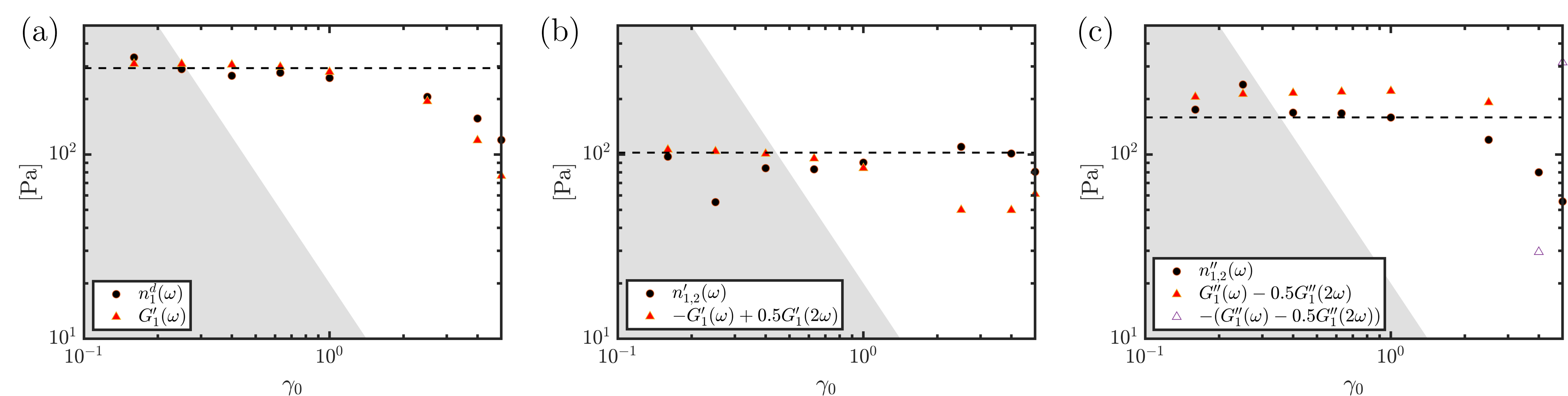}
\caption{The asymptotic low $\gamma_0$ limit for the (a) zeroth and (b), (c) second harmonics of $N_1$ signals produced in LAOS of the PDMS silicone fluid data, at an angular frequency of $\omega = 2$ rad/s ($De = \bar{\tau}_\eta \omega = 0.408$) matches the asymptotic low $\gamma_0$ limit for the shear stress data of the same material obtained at $\omega = 2$ rad/s and $2\omega = 4$ rad/s. The dotted line indicates the prediction from the 5-mode fit (\textbf{Table \ref{tab:table}}) to the linear viscoelastic spectrum. The lower limit of the force transducer is $N_{1,min} / \gamma_0^2$, which produces a line with slope $-2$ on the double logarithmic axes. Since the strain amplitude is low, as a first approximation at $De < 1$, $N_{1,min} / \gamma_0^2$ has equal contributions from $n_1^d$ and $|n_{1,2}^*|$ (as predicted by Equations \ref{UCMsmallDe_n1d} and \ref{UCMsmallDe_n12}). For the time-varying oscillatory second harmonics, a conservative estimate is to let the rheometer noise floor be the same for both $n_{1,2}'$ and $n_{1,2}''$. The area below the line with equation $20/\gamma_0^2$ in each graph is hence shaded in gray to indicate the nominal resolution of the normal force transducer.} 
\label{fig:PDMS_asymptotic}
\end{figure*}

For these moderate values of the strain amplitude $\gamma_0$, we are also able to verify that the measured $N_1$ signal is self-consistent with the shear stress signal. At each strain amplitude, the $N_1$ signal is Fourier transformed and the material functions $n_1^d (\omega)$, $n_{1,2}'(\omega)$ and $n_{1,2}''(\omega)$ were calculated from the intensity and phase angle of the zeroth and second harmonics. The shear stress signal was similarly Fourier transformed, and the material functions $G_1'(\omega)$ and $G_1''(\omega)$ were calculated from the intensity and phase of the first harmonic. As we show in \textbf{Figure \ref{fig:PDMS_asymptotic}}, the calculated material functions all shear-soften with increasing strain amplitude. More importantly, the measured $N_1$ signal is asymptotically self-consistent with the corresponding components of the shear stress signal at moderate values of the strain amplitude, as predicted by Equations \ref{consistency1} to \ref{consistency3} from second-order fluid theory. 

From theoretical considerations, we expect that applying a higher applied strain amplitude will lead to the appearance of higher harmonics in the $N_1$ signal. The experiments performed to produce \textbf{Figure \ref{fig:PDMS_asymptotic} (a)} cover a range of strain amplitudes from 0.16 to 5.0 at an angular frequency of $\omega = 2$ rad/s, corresponding to Weissenberg numbers of $Wi = 0.0653$ to 2.04. However, because measurements were performed using a CP geometry with a nonlinearly elastic material, increasing the strain amplitude to higher values ($\gamma_0 > 5$) led to sample ejection from the gap. We attempted to perform Gaborheometry strain sweeps on the material, but the magnitude of the fourth harmonic contributions to the total normal stress signal was not large enough to surpass the noise floor. These fourth harmonic terms are negligible because $\gamma_0^2 |n_{1,4}^*| \sim 10 \text{ Pa} < N_{1,min} \sim 40$ Pa. For completeness, Pipkin diagrams for the various material functions for $N_1$ in LAOS, spanning the explored parameter space of ($\omega$, $\gamma_0$), are shown in the Appendix. These are similar to those generated for the Giesekus model in \textbf{Figure \ref{fig:giesekus_contour}}. To observe the growth and behavior of higher harmonic intensities, a more strongly nonlinear fluid is required, and we now consider the response of the rheologically complex thermoplastic polyurethane (TPU) melts.

\begin{figure*}[t]
\centering
\includegraphics[width=\textwidth]{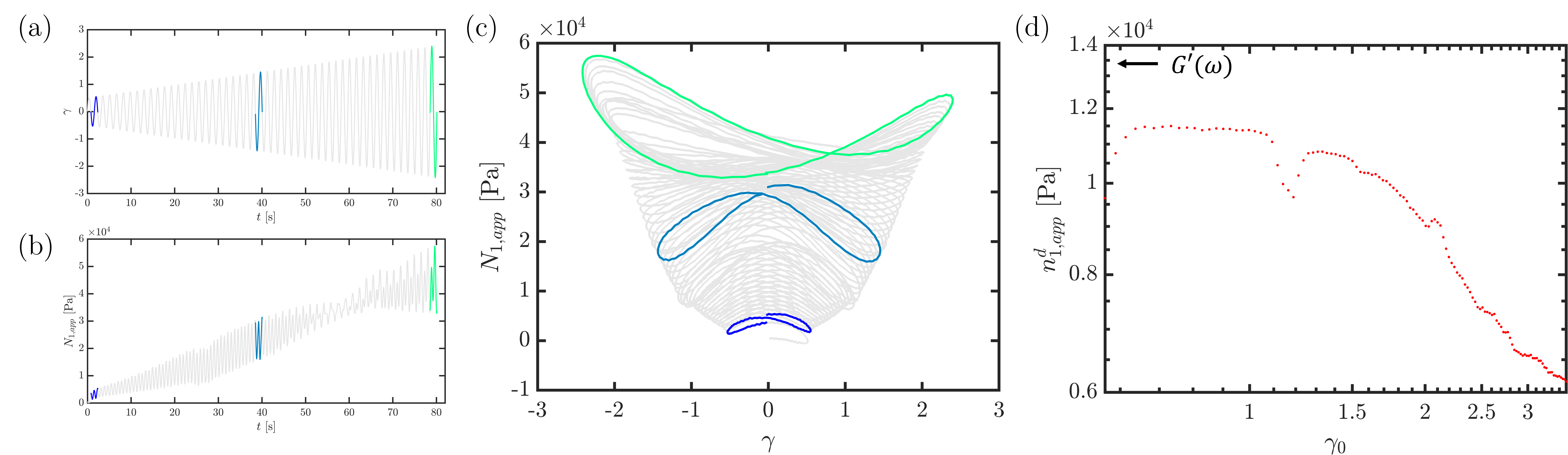}
\caption{Gaborheometry strain sweeps performed for the TPU fluid at 190$^o$C and $\omega = 4$ rad/s ($De = \bar{\tau}_\eta \omega = 5.92$). (a) An amplitude-modulated strain sweep with a linearly increasing strain amplitude acts as the input, producing (b) an oscillatory $N_{1,app}$ signal that grows in time. These two plots can be combined parametrically to obtain (c) a time-varying family of Lissajous curves for $N_{1,app}$ that increase with strain amplitude. Three representative discrete single cycles at $t_i = 2.378, 40.07, 80.12$ s (corresponding to strain amplitudes $\gamma_0(t_i) = 0.5601, 1.465, 2.426$) are highlighted in green and blue to show the progression in size and shape of the curves with time. (d) The Lissajous curves of $N_{1,app}$ are then analyzed using the windowed Gabor transform to obtain material functions for $N_{1,app}$ in LAOS at each strain amplitude applied, producing a graph of the variation of $n_{1,app}^d$ with imposed strain amplitude $\gamma_0$. The linear viscoelastic limit of $G'$ at $\omega = 4$ rad/s obtained from \textbf{Figure \ref{fig:SAOS}} and \textbf{Table \ref{tab:table}} is indicated with an arrow.} 
\label{fig:TPU_gabor}
\end{figure*}
\subsection{Pipkin Diagrams for TPU}

As was shown in \textbf{Figure \ref{fig:SAOS}}, TPU is more viscoelastic with larger dynamic shear moduli compared to PDMS, and is more rheologically complex with a broader range of relaxation mechanisms, as evidenced by the `shoulder' that becomes apparent in the van Gurp-Palmen plot. Preliminary LAOS experiments also showed that it is prone to edge fracture, and experiments have to be performed with a CPP geometry. Furthermore, at elevated temperatures, the TPU samples can degrade irreversibly when subjected to prolonged periods of shear. This necessitates the use of Gaborheometry strain sweeps to determine the material functions rapidly over a wide range of strain amplitudes. A single Gaborheometry strain sweep experiment can be performed in a few minutes, as compared to a series of discrete strain amplitude experiments, which require multiple sample loading and unloading steps and long experiment times to obtain numerous cycles of data for analysis. 

A linearly varying strain sweep was used as the input signal for eight logarithmically spaced frequencies. The frequencies were chosen to span an order of magnitude across the inverse characteristic relaxation time $\omega_c = 1/\bar{\tau}_\eta = 0.676$ rad/s as calculated from SAOS. For simplicity, a constant ramp rate $r = 0.024$ s$^{-1}$ was chosen. However, there are limits to the parameter space within the Pipkin diagram that can be explored experimentally. Below a certain strain amplitude (which depends on the frequency), the normal force signal falls below the noise floor (minimum normal force signal, $F_{min}$) of the rheometer. Furthermore, because the same ramp rate $r$ was used for all 8 sweeps, the minimum strain amplitude is constrained by the maximum Mutation number that is deemed acceptable. The maximum Mutation number $Mu_{max}$ attained during a Gabor strain sweep is typically at the beginning of the strain sweep and is calculated as $Mu_{max} = \frac{2\pi}{\omega} \frac{r}{\gamma_i}$, where $\gamma_i = \gamma_0(t=0)$ is the initial strain amplitude. As Rathinaraj and McKinley \cite{rathinaraj_gaborheometry_2023} have suggested that the Mutation number be limited to $Mu<0.1$, we set $Mu_{max} = 0.075$ for all strain sweeps, and so the initial strain amplitude for the experiments at different angular frequencies is set to be $\gamma_i = \frac{2\pi}{\omega} \frac{0.024}{0.075} = \frac{2.01}{\omega}$. On the other hand, the maximum strain amplitude accessible is limited by the onset of edge fracture in the sample, which is identified by an abrupt change in the axial force being measured.

For the CPP geometry, the axial force measured by the force transducer of the rheometer is not the true $N_1$ signal. Instead, it is an apparent $N_1$ signal given by the following expression \cite{meissner_measuring_1989,schweizer_measurement_2002}: 
\begin{equation} \label{cpp}
\begin{split}
    N_{1,app} &= N_1 + 2(N_1 + 2N_2) \ln \left(\frac{R}{R_{stem}} \right) \\
    &= N_1 \left(1 + 2 \left( 1 + \frac{2 N_2}{N_1} \right) \ln \left(\frac{R}{R_{stem}} \right) \right)
\end{split}
\end{equation}

The second term in equation \ref{cpp} can be viewed as a correction term to $N_1$ due to the use of the CPP geometry as compared to a CP geometry. The apparent $N_1$ signal is thus influenced by the second normal stress difference $N_2$, and it is difficult to decouple its contribution from the true $N_1$, because the ratio of $N_2 / N_1$ is also a function of strain amplitude $\gamma_0$ and angular frequency $\omega$. For example, for the Giesekus model, the ratio of the respective means of $N_2/N_1$ at a fixed angular frequency decreases with increasing $\gamma_0$. \cite{bird_dynamics_1987} In principle, it is possible to take another set of measurements with a different geometry (with a different value of $R/R_{stem}$) to remove the contribution from $N_2$ by subtraction. \cite{costanzo_review_2024} This requires additional time-consuming experiments and is sensitive to error amplification due to the subtraction of two terms of similar magnitude. To illustrate the general utility of the Fourier-Tschebyshev framework for $N_1$ and the Gabor strain sweep process, the apparent $N_1$ signals were analyzed as is without modification. Furthermore, because both $N_1$ and $N_2$ are even functions of strain amplitude, the apparent $N_1$ is also an even function of the strain amplitude and can be expressed using the Fourier-Tschebyshev formulation for $N_1$. For instance, $n_{1,app}^d$ is similarly defined as the zeroth harmonic term of the time-varying apparent $N_1$ signal divided by the square of the strain amplitude, $\frac{N_{1,app}}{\gamma_0^2}$.

\begin{figure*}
\centering
\includegraphics[width=\textwidth]{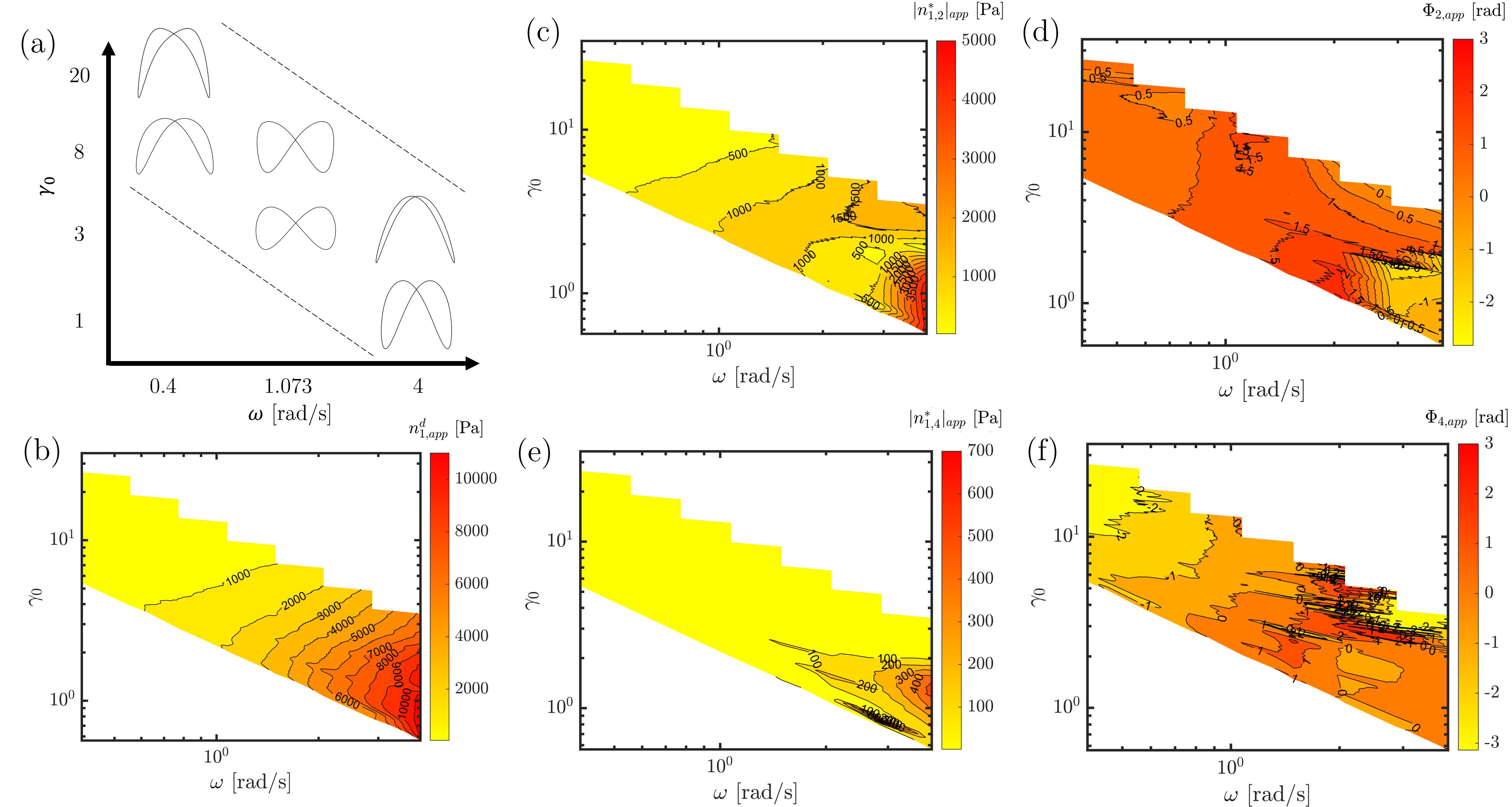}
\caption{Pipkin diagrams of the material functions of the apparent $N_1$ signal in LAOS for the TPU melt at 190$^o$C. The Deborah number ranges from $De = \bar{\tau}_\eta \omega = 0.592$ to 5.92. (a) Representative shapes of the Lissajous curves of apparent $N_1$ at selected strain amplitudes and angular frequencies. The evolution in the magnitude of the DC coefficient $n_{1,app}^d (\gamma_0, \omega)$ with strain amplitude and angular frequency is shown in (b) and then the strength and phase information for the second and fourth harmonics using the notation of equation \ref{notation} are shown in (c), (d) and (e), (f) respectively.} 
\label{fig:TPU_contour}
\end{figure*}

As we show in \textbf{Figure \ref{fig:TPU_gabor}}, similar to the Giesekus model, a Gabor strain sweep performed with a linearly increasing strain amplitude leads to a steadily increasing apparent $N_1$ signal, which can also be viewed as a series of Lissajous curves that grow in magnitude with time. These curves were analyzed using the windowed Gabor transform to obtain the relevant material functions, such as $n_{1,app}^d$ at each time instance (or strain amplitude). For the true $N_1$ signal, we expect $n_1^d (\omega) = G' (\omega)$ from second-order fluid theory, and this is indicated by an arrow on the ordinate of \textbf{Figure \ref{fig:TPU_gabor} (d)}. Since what we measure is instead an apparent $n_1^d$ that is modified by a contribution from the second normal stress difference, $n_{1,app}^d$ plateaus to an asymptotic value different from $G'(\omega)$. Because $N_2/N_1 <0$ for most polymeric systems \cite{maklad_review_2021}, we can anticipate that this will also be the case for the TPU sample. From the data shown in \textbf{Figure \ref{fig:TPU_gabor} (d)}, and using the values of $R$ and $R_{stem}$ from Section IV.A, we estimate that for this TPU, $N_{2,0} / N_{1,0} = \Psi_{2,0} / \Psi_{1,0} \approx -0.54$. Although this ratio is larger than expected for entangled melts of linear homopolymers, for liquid crystal-like systems, which may be characteristic of the hard block domains within TPU, experimental measurements of concentrated solutions of rod-like PBLG (poly($\gamma$-benzyl L-glutamate)) in \textit{m}-cresol (with concentration > 12.3
wt\%) \cite{magda_shear_1991} indicate that $-N_2 / N_1 > 0.5$. It is also generally acknowledged that large values of $N_2$ strongly drive secondary flows, and edge fracture is highly likely.\cite{maklad_review_2021} This is indeed the case for this TPU melt, and this further rationalizes the estimate of $N_{2,0} / N_{1,0}$, illustrating the utility of a CPP geometry in measuring normal stress differences in a highly elastic polymer melt, and obtaining estimates of $N_{2,0} / N_{1,0}$.

The Pipkin diagrams obtained from the Gabor strain sweeps are shown in \textbf{Figure \ref{fig:TPU_contour}}. In \textbf{Figure \ref{fig:TPU_contour} (a)}, we show just a few representative examples of the Lissajous orbits at three of the eight test frequencies considered. These were reconstructed from the Fourier-Tschebyshev analysis of the experimental $N_{1,app} (t)$ signal using the DC, second order, and fourth order terms, which are then plotted individually in \textbf{Figures \ref{fig:TPU_contour}(b)} to \textbf{(f)}. We observe that an evolution in the various harmonics of $N_{1,app}$ occurs with strain amplitude. Similar to the Giesekus model, the values of the contour lines for the material functions $n_{1,app}^d$ and $|n_{1,2}^*|_{app}$ progressively increase from left to right at low $\gamma_0$. However, apart from that similarity, the general features of the material property surfaces are clearly more complex than those that can be predicted by the single quadratic nonlinearity of the Giesekus model. In particular, the contour plots of the fourth harmonic intensity and phase angle, which provide a measure of the onset of full material nonlinearity, are very different compared to the predictions of the simple Giesekus model. The fourth order terms are clearly above the minimum $N_1$ signal of 40 Pa, but are also much smaller than the second harmonic terms. There is a sign change in $\Phi_2$ and $\Phi_4$ at separate critical angular frequencies, which is also predicted by the Giesekus model (refer to \textbf{Figure 12} in the Appendix). However, the locus of the boundary at which $\Phi_2 = 0$ and $\Phi_4 = 0$ (as indicated by the local clustering of black contours where variation in the magnitude is rapid) is different compared to the predictions of a single mode Giesekus model. This indicates that, despite the widespread utility of the Giesekus model as a good first approximation to the nonlinear shear rheology of polymer melts \cite{larson_constitutive_1988}, it does not universally describe the nonlinear behavior of a more rheologically complex material such as a thermoplastic polyurethane melt. Correspondingly, this motivates the future use of LAOS extended fingerprints to distinguish between different nonlinear model predictions or to discover new constitutive models using machine learning tools \cite{mangal_data-driven_2025}. 

\section{Conclusion and Discussion}

We have developed a Fourier-Tschebyshev framework for systematically representing the material functions that characterize the evolution in $N_1$ with frequency and strain amplitude in LAOS. We hope this framework will be adopted to compactly and quantitatively represent the $N_1$ material response for complex fluids in LAOS as prevalently as the elastic and viscous Tschebyshev coefficients for the shear stress in LAOS. We have shown that the use of slow strain sweeps and the time-localization properties of Gaborheometry can be extended to extract these new material functions that quantitatively describe the evolution of $N_1$ in LAOS as a function of strain amplitude, producing rich datasets spanning multiple strain amplitudes within a single experiment and revealing the appearance and growth of higher harmonics at high strain amplitudes. On the other hand, we also showed that self-consistency between the measured values of $N_1$ in LAOS and the oscillating shear stress at low to moderate strain amplitudes is satisfied for real fluids and can act as a useful check of the torque and normal force transducer calibration. The predicted offset between the values of $n_{1,app}^d$ and $G'$ measured at moderate strain amplitudes using the CPP geometry can also be used to provide an estimate of $N_{2,0} / N_{1,0}$. 

These complex feature representations for $N_1 (t; \omega, \gamma_0)$ complement the information gained from analyzing the spectral composition of oscillatory shear stress measurements in LAOS. They are not only helpful in understanding the nonlinear viscoelastic material response to an applied strain in terms of the underlying physics, but also serve as enhanced rheological fingerprints that more fully characterize a complex fluid. For example, the recently developed Rheological Universal Differential Equations (RUDEs) framework \cite{lennon_scientific_2023} leveraged LAOS shear stress datasets to learn the form of constitutive relations based on a generalized Upper Convected Maxwell model. The major advantage of the universal differential equation approach is the preservation of full material frame indifference through the architecture of the tensor basis neural network. Expanding this training dataset to include $N_1$ data in LAOS, or even apparent $N_1$ data, may enable the RUDE to learn more accurate feature representations for an unknown nonlinear viscoelastic fluid of interest. We anticipate that in the future this will enable the production of accurate frame-indifferent tensorial models across different flow protocols, creating complete "digital fluid twins" that can reliably be utilized beyond the viscometric flows associated with "digital twin rheometers". Future work will focus on reproducing the strongly nonlinear rheological behavior of TPUs in LAOS and other strong shearing deformations through the discovery of an appropriate constitutive equation using the RUDEs framework. 

\begin{acknowledgments}
The authors acknowledge the financial support of the Lubrizol Corporation. N.K. also acknowledges conference grant support from the National University of Singapore Overseas Graduate Scholarship (NUS-OGS). 
\end{acknowledgments}

\section*{Conflict of Interest}
The authors have no conflicts to disclose.

\section*{Data Availability Statement}
The data that support the findings of this study are available from the corresponding author upon reasonable request.

\appendix

\section{Medium Amplitude Oscillatory Shear (MAOS) of the Giesekus Model}

Gurnon and Wagner \cite{gurnon_large_2012} performed a MAOS expansion of the Giesekus model to find the nonlinear corrections to the oscillating shear stress which appear at third-order in strain. The fourth-order terms for the oscillatory first normal stress difference were found by applying a similar expansion for only even $n$:
\begin{widetext}
\begin{equation}
    N_1 (t; \gamma_0, \omega) = \sum_{\substack{j=2 \\ \text{even}}}^\infty \sum_{\substack{n=0 \\ \text{even}}}^j \left( (A_{11,n}^{(j)} (\gamma_0, \omega) - A_{22,n}^{(j)} (\gamma_0, \omega)) \sin (n \omega t) + (B_{11,n}^{(j)} (\gamma_0, \omega) - B_{22,n}^{(j)} (\gamma_0, \omega)) \cos (n \omega t) \right)
\end{equation}
\end{widetext}

Using the same expansion technique, the terms contributing to the first normal stress difference at each order $(j)$ are:
\begin{widetext}
\begin{equation}
\begin{split}
    N_1 (t; \gamma_0, \omega) &= (B_{11,0}^{(2)} - B_{22,0}^{(2)}) + (A_{11,2}^{(2)} - A_{22,2}^{(2)}) \sin (2\omega t) + (B_{11,2}^{(2)} - B_{22,2}^{(2)}) \cos (2\omega t) \\
    &+ (B_{11,0}^{(4)} - B_{22,0}^{(4)}) + (A_{11,2}^{(4)} - A_{22,2}^{(4)}) \sin (2\omega t) + (B_{11,2}^{(4)} - B_{22,2}^{(4)}) \cos (2\omega t) \\
    &+ (A_{11,4}^{(4)} - A_{22,4}^{(4)}) \sin (4\omega t) + (B_{11,4}^{(4)} - B_{22,4}^{(4)}) \cos (4\omega t) \\
    &+ \mathcal{O} (\gamma_0^6, \omega^6)
\end{split}
\end{equation}
\end{widetext}

Each of the coefficients above was found by substituting the expression for the shear and normal stress into the coupled system of differential equations for the stress components of the Giesekus model. The values of the second order terms $B_{11,0}^{(2)}$, $B_{22,0}^{(2)}$, $A_{11,2}^{(2)}$, $B_{11,2}^{(2)}$, $A_{22,2}^{(2)}$ and $B_{22,2}^{(2)}$ were previously found by Gurnon and Wagner \cite{gurnon_large_2012} (Equations 16 to 21 in their paper) and will not be repeated here. After a great deal of tedious mathematics, the following expressions were found for the fourth order corrections: 
\begin{widetext}

\begin{equation}
    B_{11,0}^{(4)} = -\frac{\alpha G \gamma_0^4 De^4 (30 + 70 De^2 + 
       16 De^4 + 
       \alpha^2 (15 + 35 De^2 + 8 De^4) - 
       2 \alpha (21 + 47 De^2 + 
          8 De^4))}{8 (1 + De^2)^3  (1 + 
       4 De^2)}
\end{equation}
\begin{equation}
    A_{11,2}^{(4)} = -\frac{\alpha G \gamma_0^4 De^5 (53+381 De^2 + 567 De^4+ 59 De^6 - 180 De^8 + x_1)}{2 (1 + De^2)^4 (1 + 4 De^2)^2 (1+9 De^2)}
\end{equation}
\begin{equation}
\begin{split}
    x_1 &= - 4 \alpha^2 (-8-40 De^2 + 37 De^4 + 141 De^6) \\ 
    &+ \alpha (-82-496 De^2 - 230 De^4 + 760 De^6 + 288 De^8)
\end{split}
\end{equation}
\begin{equation}
    B_{11,2}^{(4)} = \frac{\alpha G \gamma_0^4 De^4 (\alpha^2 (-5 + 35 De^2 + 421 De^4 + 489 De^6 - 180 De^8) + x_2)}{2 (1 + De^2)^4 (1 + 4 De^2)^2 (1+9 De^2)}
\end{equation}
\begin{equation}
\begin{split}
    x_2 &= - 2 \alpha (-7 + 25 De^2 + 461 De^4 + 777 De^6 + 204 De^8) \\ 
    &+ 2 (-5 + De^2 + 234 De^4 + 553 De^6 + 361 De^8 + 36 De^{10})
\end{split}
\end{equation}
\begin{equation}
    A_{11,4}^{(4)} = \frac{\alpha G \gamma_0^4 De^5 (-(1 + De^2)^2 (53 - 71 De^2 - 2068 De^4 +        576 De^6) + x_3)}{4 (1 + De^2)^4 (1 + 4 De^2)^2 (1 +   25 De^2 + 144 De^4)}
\end{equation}
\begin{equation}
\begin{split}
    x_3 &= 4 \alpha^2 (-8 + 72 De^2 + 387 De^4 - 707 De^6 + 66 De^8) \\
    &+ \alpha (82 - 332 De^2 - 4206 De^4 - 608 De^6 + 3184 De^8)
\end{split}
\end{equation}
\begin{equation}
    B_{11,4}^{(4)} = \frac{\alpha G \gamma_0^4 De^4 (-2  (1 + De^2)^2 (5 - 179 De^2 - 880 De^4 +      1824 De^6) + x_4)}{8 (1 + De^2)^4 (1 + 4 De^2)^2 (1 + 25 De^2 + 144 De^4)}
\end{equation}
\begin{equation}
\begin{split}
    x_4 &= \alpha^2 (-5 + 305 De^2 + 237 De^4 - 6005 De^6 + 2708 De^8) \\ 
    &- 2 \alpha (-7 + 329 De^2 + 1275 De^4 - 4177 De^6 - 4348 De^8 + 768 De^{10})
\end{split}
\end{equation}

\begin{equation}
    B_{22,0}^{(4)} = -\frac{\alpha^2 G \gamma_0^4 De^4 (-2 (9 + 23 De^2 + 
       8 De^4) + 
    \alpha (15 + 35 De^2 + 8 De^4))}{
 8 (1 + De^2)^3 (1 + 4 De^2)}
\end{equation}

\begin{equation}
    A_{22,2}^{(4)} = \frac{\alpha^2 G \gamma_0^4 De^5 (18 + 107 De^2 + 
     10 De^4 - 259 De^6 - 
     36 De^8 + 
     x_5)}{(1 + De^2)^4 (1 + 
     4 De^2)^2 (1 + 9 De^2)}
\end{equation}
\begin{equation}
    x_5 = 2 \alpha (-8 - 40 De^2 + 37 De^4 + 
        141 De^6)
\end{equation}

\begin{equation}
    B_{22,2}^{(4)} = -\frac{\alpha^2 G \gamma_0^4 De^4 (-6 + 26 De^2 + 
       442 De^4 + 698 De^6 + 
       x_6)}{2 (1 + 
       De^2)^4 (1 + 4 De^2)^2 (1 + 
       9 De^2)}
\end{equation}
\begin{equation}
    x_6 = \alpha (5 - 35 De^2 - 421 De^4 - 
          489 De^6 + 180 De^8)
\end{equation}

\begin{equation}
    A_{22,4}^{(4)} = \frac{\alpha^2 G \gamma_0^4 De^5 (
     3 (6 - 35 De^2 - 252 De^4 - 
        15 De^6 + 196 De^8) + x_7)}{2 (1 + 
     De^2)^4 (1 + 4 De^2)^2 (1 + 
     25 De^2 + 144 De^4)}
\end{equation}
\begin{equation}
    x_7 = 2 \alpha (-8 + 
        72 De^2 + 387 De^4 - 
        707 De^6 + 66 De^8)
\end{equation}

\begin{equation}
    B_{22,4}^{(4)} = \frac{\alpha^2 G \gamma_0^4 De^4 (- 
     6 (-1 + 51 De^2 + 117 De^4 - 
        547 De^6 - 516 De^8 + 
        96 De^{10}) + x_8)}{8 (1 + De^2)^4 (1 + 
     4 De^2)^2 (1 + 25 De^2 + 
     144 De^4)}
\end{equation}
\begin{equation}
    x_8 = \alpha (-5 + 
        305 De^2 + 237 De^4 - 
        6005 De^6 + 2708 De^8)
\end{equation}

\end{widetext}

The following expressions are the leading order harmonics:
\begin{widetext}
\begin{equation}
    B_{11,0}^{(2)} - B_{22,0}^{(2)} = \frac{G \gamma_0^2 De^2}{1 + De^2} = \gamma_0^2 n_1^d (\omega)
\end{equation}

\begin{equation}
    A_{11,2}^{(2)} - A_{22,2}^{(2)} = \frac{3 G \gamma_0^2 De^3}{(1 + De^2)(1 + 4 De^2)} = \gamma_0^2 n_1'' (\omega)
\end{equation}

\begin{equation}
    B_{11,2}^{(2)} - B_{22,2}^{(2)} = \frac{G \gamma_0^2 De^2 (1 - 2 De^2)}{(1 + De^2)(1 + 4 De^2)} = \gamma_0^2 n_1' (\omega)
\end{equation}

\begin{equation}
    A_{11,4}^{(4)} - A_{22,4}^{(4)} = \frac{\alpha G \gamma_0^4 De^5 (-53 + 18 De^2 + 
     2139 De^4 + 1492 De^6 - 
     576 De^8 + 
     x_9 )}{4 (1 + De^2)^3 (1 + 
     4 De^2)^2 (1 + 25 De^2 + 
     144 De^4)}
\end{equation}
\begin{equation}
    x_9 = 2 \alpha (23 - 84 De^2 - 1263 De^4 + 
        1004 De^6)
\end{equation}

\begin{equation}
    B_{11,4}^{(4)} - B_{22,4}^{(4)} = - \frac{\alpha G \gamma_0^4 De^4 (5 - 174 De^2 - 
       1059 De^4 + 944 De^6 + 
       1824 De^8 + 
       x_{10})}{4 (1 + 
       De^2)^3 (1 + 4 De^2)^2 (1 + 
       25 De^2 + 144 De^4)}
\end{equation}
\begin{equation}
    x_{10} = 4 \alpha (-1 + 45 De^2 + 186 De^4 - 
          820 De^6 + 120 De^8)
\end{equation}
\end{widetext}

Note that if $\alpha = 0$, the fourth order terms vanish and the Upper Convected Maxwell model (with only the zeroth and second harmonics) is recovered. In the limit of $\gamma_0 \ll 1$, the following relation should hold:
\begin{equation}
    |n_{1,4}^*| = \frac{\sqrt{(A_{11,4}^{(4)} - A_{22,4}^{(4)})^2 + (B_{11,4}^{(4)} - B_{22,4}^{(4)})^2}}{\gamma_0^2} \propto \gamma_0^2
\end{equation}

\begin{figure}
\centering
\includegraphics[width=0.4\textwidth]{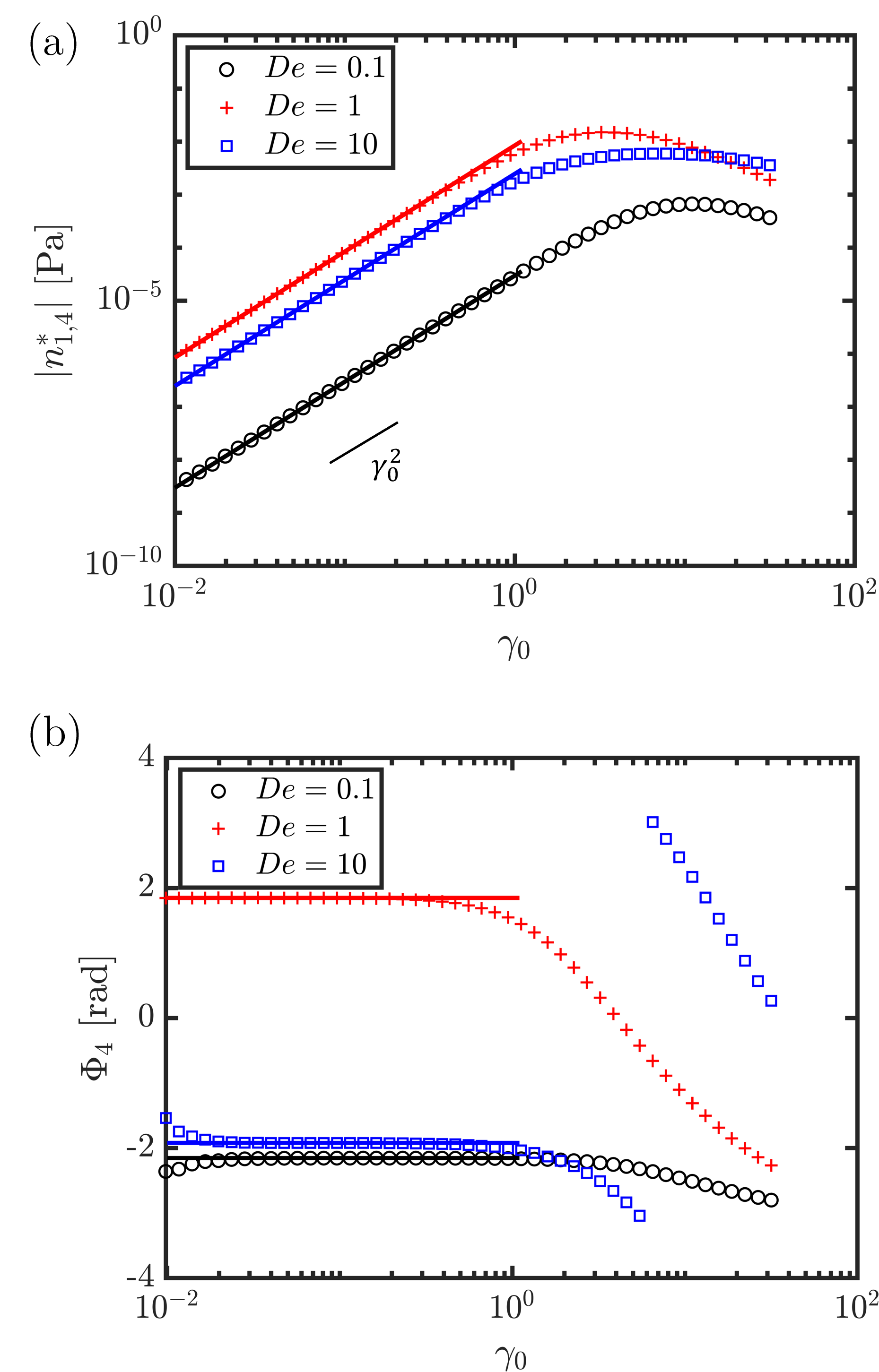}
\caption{Verification of MAOS analytic solutions for $N_1$ (solid lines) against the full numerical simulations (symbols) at various Deborah numbers for the Giesekus model ($G=1$ Pa, $\tau = 1$ s, $\alpha = 0.4$). The Deborah numbers were varied by changing the imposed angular frequency $\omega$. At $De = 10$, there is a sign change in the phase angle $\Phi_4$ at $\gamma_0 \approx 6$.} 
\label{fig:giesekus_MAOS}
\end{figure}

The above expression is plotted as a line in \textbf{Figure \ref{fig:giesekus_MAOS}} where the slope is indeed found to be 2 for $\gamma_0 \ll 1$. The symbols in \textbf{Figure \ref{fig:giesekus_MAOS}} are the parameters obtained from Fourier-Tschebyshev analysis of the Giesekus model solved at each value of ($\omega$, $\gamma_0$). The satisfying agreement between the analytical asymptotic solutions and the numerical calculations gives confidence that the extraction of harmonics by Fourier-Tschebyshev analysis of time-series data is accurate. The slight deviations observed for the phase angle at very low $\gamma_0$ are due to numerical integration error from the stiffness of the system of differential equations.

\section{PDMS Contour Plots}

\begin{figure*}
\centering
\includegraphics[width=0.8\textwidth]{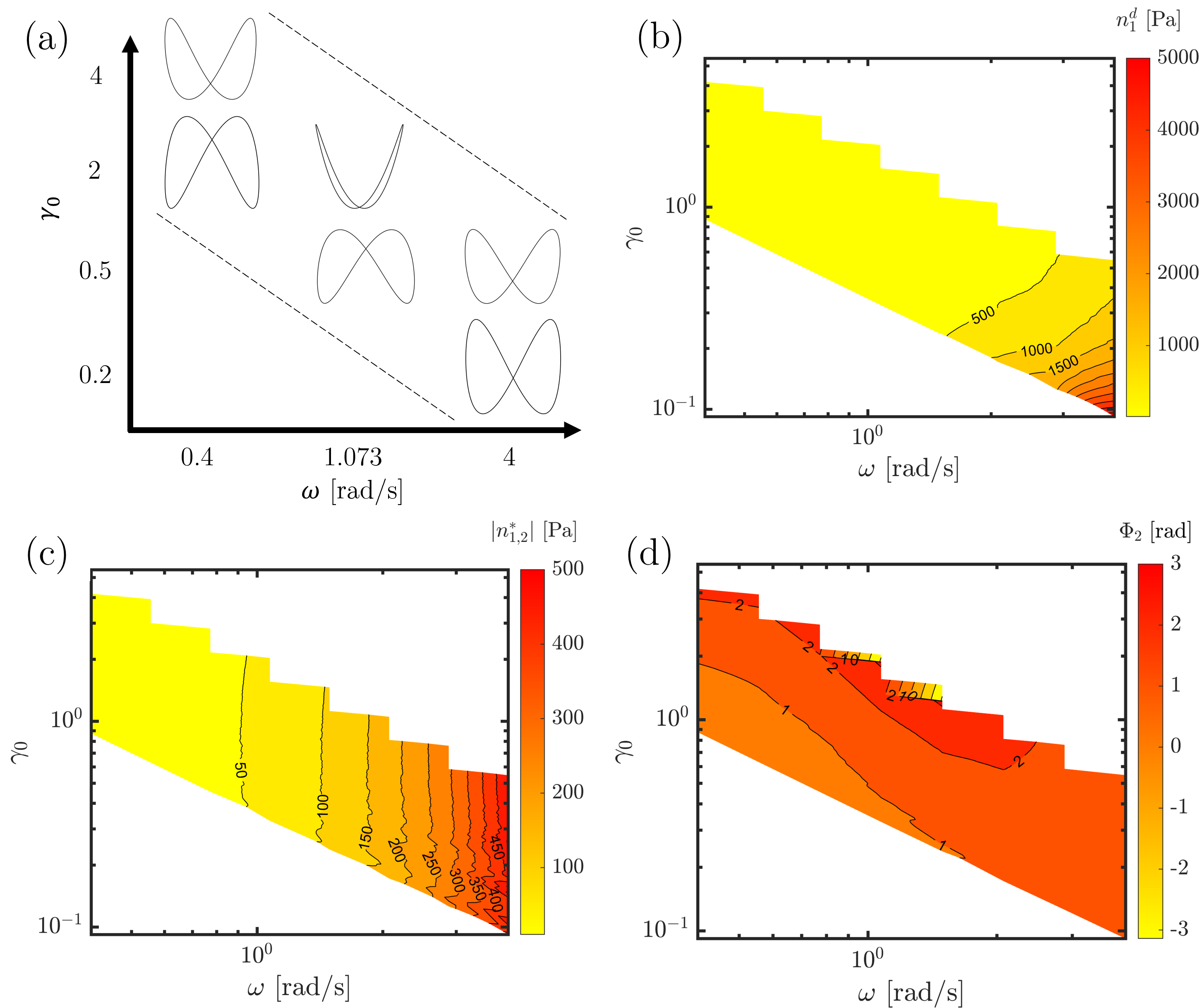}
\caption{Pipkin diagrams showing the evolution in the $N_1$ LAOS coefficients measured in the silicone fluid. (a) Representative shapes of the Lissajous curves of apparent $N_1$ at selected strain amplitudes and angular frequencies. The magnitude of the DC coefficient $n_1^d$ is shown in (b) and then the strength and phase information for the second harmonic is shown in (c) and (d). The intensity of the fourth harmonic was too weak and hence the Pipkin diagrams of the fourth harmonic were not plotted.} 
\label{fig:PDMS_contour}
\end{figure*}

The contour plots for the PDMS silicone fluid are shown in \textbf{Figure \ref{fig:PDMS_contour}}, where Gaborheometry strain sweeps for eight logarithmically spaced frequencies from 0.4 to 4 rad/s, for a constant ramp rate of 0.005 s$^{-1}$ (maximum Mutation number $Mu_{max} = 0.1$), were applied. Because a cone and plate (CP) geometry was used to obtain the true $N_1$ signal, edge fracture occurs at lower values of $\gamma_0$ compared to the TPU fluid. Hence, experiments for the PDMS were carried out at a larger maximum Mutation number and a lower ramp rate compared to the TPU sample. We observe that the values of $|n_{1,2}^*|$ generally remain near the value of $n_1^d$. This consistent with the prediction from second-order fluid theory that $n_1^d \sim |n_{1,2}^*|$ (Equations \ref{UCMlargeDe_n1d} and \ref{UCMlargeDe_n12}), which is valid here since the Deborah number ranges from $De = \bar{\tau}_\eta \omega = 0.0816$ to 0.816.

When working with experimental data, we also observe more strongly the trade-off between time and frequency resolution. At low $\gamma_0$, the data is noisier as it is closer to the noise floor of the force transducer of the rheometer. Because the values of $n_1^d$ and $|n_{1,2}^*|$ are obtained by dividing the (noisy) measured $N_1$ signal by $\gamma_0^2$, at small $\gamma_0$, such errors are magnified substantially. 

\section{TPU Spectrograms}

\begin{figure}
\centering
\includegraphics[width=0.4\textwidth]{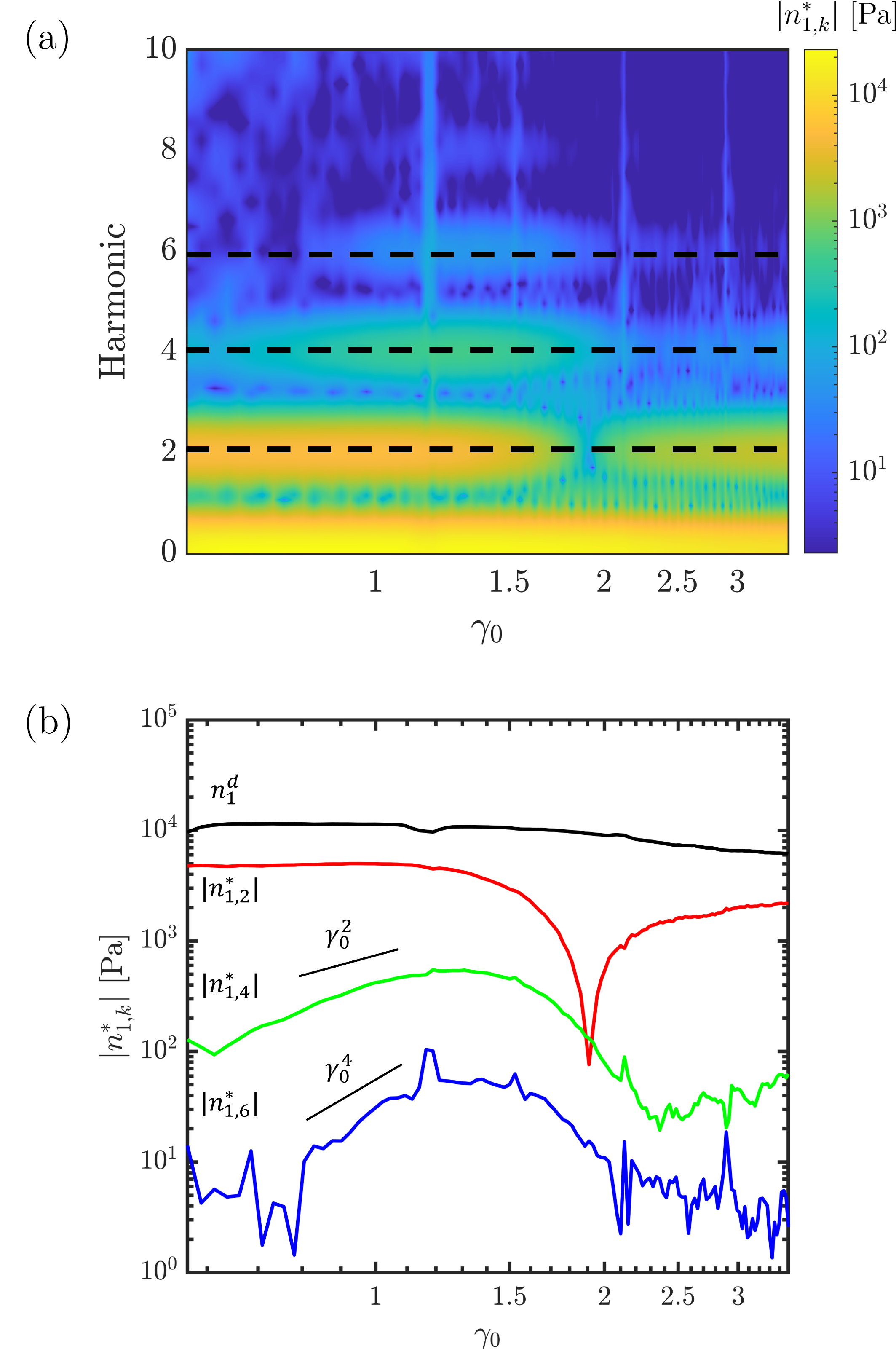}
\caption{Gabor spectrograms of experimental TPU data obtained using a slowly ramped strain amplitude ($\gamma_0(t)=0.503+0.024t$) at 190$^o$C for a frequency $\omega =4$ rad/s. The leading order nonlinearity at low strain amplitudes for each harmonic intensity varies as an even multiple of $\gamma_0$, as is expected from the MAOS regime. Harmonics of higher order than the 6th are much weaker in intensity.} 
\label{fig:TPU_spectrogram}
\end{figure}

Gabor spectrograms are another way of showing the intensities of all harmonics inherent in a signal. Each of the contour plots shown previously only traces the evolution of the harmonic of interest as the strain amplitude $\gamma_0$ increases. As shown in \textbf{Figure \ref{fig:TPU_spectrogram}}, the intensities of the odd harmonics are small compared to those of the even harmonics. In particular, the intensity of the first harmonic is less than 10\% of the intensity of the zeroth or the second harmonic. In the context of shear stress LAOS signals, it has been demonstrated that even harmonics may arise due to wall slip \cite{graham_wall_1995} or inertia \cite{atalik_occurrence_2004}. The same reasons may explain the presence of odd harmonics in the $N_1$ signal, and the intensities of these harmonics can be viewed as a proxy for the fidelity of the data. Large intensities of odd harmonics in the $N_1$ signal (which should theoretically only contain even harmonics) would indicate that experimental artifacts may be corrupting the data. 

\renewcommand{\refname}{REFERENCES}

% \nocite{*}
\bibliography{LAOSandnormalstresses}% Produces the bibliography via BibTeX.
% \bibliography{aipsamp}% Produces the bibliography via BibTeX.

\end{document}